\documentclass[a4paper,twocolumn,notitlepage,prx,superscriptaddress]{revtex4-2}
\pdfoutput=1
\usepackage[english]{babel}
\usepackage{graphicx}
\usepackage{amsmath}
\usepackage{amsfonts}
\usepackage{amssymb}
\usepackage{xcolor}
\usepackage{physics}
\usepackage{bbm}
\usepackage{academicons}
\definecolor{orcidlogocol}{HTML}{A6CE39}
\usepackage{dsfont}
\usepackage{hyperref}
\hypersetup{
    bookmarks=false,         
    unicode=false,          
    pdftoolbar=false,        
    pdfmenubar=true,        
    pdffitwindow=false,     
    pdfstartview={FitH},    
    pdftitle={Many-body localization proximity effect in two-species bosonic Hubbard model},    
    pdfauthor={Brighi et al},     
    pdfsubject={},   
    pdfcreator={},   
    pdfproducer={}, 
    pdfkeywords={disorder, quantum transport, ergodicity breaking},
    pdfnewwindow=true,      
    colorlinks=true,       
    linkcolor=black,          
    citecolor=blue,        
    filecolor=magenta,      
    urlcolor=blue           
}
\usepackage[normalem]{ulem}
\newcommand{\blubullet}{{\color{blue}\bullet}}
\newcommand{\redbullet}{{\color{red}\bullet}}
\newcommand{\orcidicon}{\includegraphics[width=.0125\columnwidth]{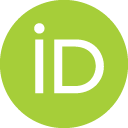}}
\newcommand{\orcid}[1]{\href{https://orcid.org/#1}{\textcolor{orcidlogocol}\orcidicon}}

\begin{document}
\title{
Many-body localization proximity effect in two-species bosonic Hubbard model
}
\author{Pietro Brighi\orcid{0000-0002-7969-2729}}
\affiliation{Institute of Science and Technology Austria (ISTA), Am Campus 1, 3400 Klosterneuburg, Austria}
\author{Marko Ljubotina\orcid{0000-0003-0038-7068}}
\affiliation{Institute of Science and Technology Austria (ISTA), Am Campus 1, 3400 Klosterneuburg, Austria}
\author{Dmitry A. Abanin\orcid{0000-0002-2461-0271}}
\affiliation{Department of Theoretical Physics, University of Geneva, 24 quai Ernest-Ansermet, 1211 Geneva, Switzerland}
\author{Maksym Serbyn\orcid{0000-0002-2399-5827}}
\affiliation{Institute of Science and Technology Austria (ISTA), Am Campus 1, 3400 Klosterneuburg, Austria}

\date{\today}
\begin{abstract}
The many-body localization (MBL) proximity effect is an intriguing phenomenon where a thermal bath localizes due to the interaction with a disordered system. 
The interplay of thermal and non-ergodic behavior in these systems gives rise to a rich phase diagram, whose exploration is an active field of research.
In this work, we study a bosonic Hubbard model featuring two particle species representing the bath and the disordered system.
Using state of the art numerical techniques, we investigate the dynamics of the model in different regimes, based on which we obtain a tentative phase diagram as a function of coupling strength and bath size.
When the bath is composed of a single particle, we observe clear signatures of a transition from an MBL proximity effect to a delocalized phase.
Increasing the bath size, however, its thermalizing effect becomes stronger and eventually the whole system delocalizes in the range of moderate interaction strengths studied.
In this regime, we characterize particle transport, revealing diffusive behavior of the originally localized bosons.
\end{abstract}
\maketitle

\section{Introduction}

The concept of a thermal bath plays a central role in the description of equilibrium systems in statistical mechanics.
Typically, a bath is defined as a large system, unaffected by the coupling to the system considered, whose role is to provide a reservoir of energy to the system and thermalize it.
Recent advances in the field of quantum simulators, however, allow for an unprecedented degree of control over the experimental setup parameters. 
This has introduced the possibility of studying small quantum baths and their interactions with otherwise isolated quantum systems~\cite{Rubio-Abadal2019,Leonard2020}.
In this scenario one can also study the effect of the coupling to the system on the bath itself.

Intuitively, coupling an ergodic system to a bath will result in a combined system with similar properties.
However, in the cases where the system is non-ergodic and hence does not satisfy the eigenstate thermalization hypothesis~\cite{Deutsch1991,Srednicki1994}, more exotic phenomena can be observed.
A natural question then arises regarding the various phases in these systems and their stability with respect to the model parameters.
An example of non-ergodic systems are integrable models~\cite{Takahashi1999,Rigol2007,Rigol2008}, however these are known to be unstable to weak perturbations~\cite{D'Alessio2016}.
Thus the expected outcome is the eventual thermalization of the system through the coupling with the bath, although recent studies have shown that in special cases seemingly stable bound states can form~\cite{Roy2019,Ljubotina2022a}.
A more robust scenario is offered by disordered systems, since the many-body localized (MBL) phase~\cite{Basko2006,Gornyi2005a,Huse-rev,Abanin2019} arising there represents an example of strong ergodicity breaking stable to weak perturbations, thus providing an interesting case of study for the fate of non-ergodic systems coupled to baths.

In this scenario, two distinct outcomes are possible. 
First, similarly to the case of integrable models, the coupling to the bath could lead to the system thermalizing. 
Alternatively, the quantum bath can be affected by the coupling to the disordered system leading to a breakdown of thermalization in the bath itself, via the so-called  MBL proximity effect~\cite{Nandkishore2015a}.
Motivated by experiments~\cite{Rubio-Abadal2019,Leonard2020} and by the fundamental question of the stability of MBL in the presence of a bath, a large number of works~\cite{Huse2015,Nandkishore2015a,Knap2017,Hyatt2017,Luitz2017,Pollmann2019,Goihl2019,Bera2021,Krause2021,BarLev2022,Brighi2022a,Brighi2022b} considered the interplay of disordered and ergodic systems in different setups.
Of these, Refs.~\cite{Huse2015,Knap2017,Luitz2017,Pollmann2019,Bera2021,BarLev2022} studied the case when the back-action of the MBL system onto the bath can be discarded, thus excluding the possibility of an MBL proximity effect. 
A different setting, where the thermal bath is modeled as an ergodic many-body system, thus being potentially prone to localization due to the interaction with the MBL system, was considered by Refs.~\cite{Nandkishore2015a,Hyatt2017,Brighi2022a,Brighi2022b}. 

In this context, some recent studies~\cite{Brighi2022a,Brighi2022b} provided numerical evidence for the stability of the MBL proximity effect in the special case of a quantum bath consisting of a single particle. 
The bath-MBL coupling was realized using a Hubbard model with two hard-core bosonic species, inspired by the experimental setup of Ref.~\cite{Rubio-Abadal2019}.
In this model, one of the particle types, the \textit{disordered bosons} experience a random on-site potential, thus representing a non-ergodic localized system.
The second species -- the \textit{clean bosons} -- are not subject to the random on-site potential, and play the role of a quantum bath of variable size depending on the number of such particles.
Using matrix product states (MPS)~\cite{Verstraete2006} based algorithms for numerical time evolution and for accessing highly excited eigenstates, Refs.~\cite{Brighi2022a,Brighi2022b} demonstrated that at strong disorder and strong bath-system coupling the single clean boson  fails to thermalize an extensive number of disordered particles and gets localized by the disorder induced by the interaction with the localized particles. 
Despite providing evidence of the realization of the MBL proximity effect, Refs.~\cite{Brighi2022a,Brighi2022b} left many open questions. 
In particular, the fate of the system at weaker interactions and in the case of larger baths (for instance, at a finite density of clean particles) remained unexplored. 
Moreover, a recent work~\cite{Sierant2022} challenged the conclusions on the stability of localization, hinting at the possibility of delocalization at longer times. 

In this work we use state of the art numerical methods with large computational resources to consider hitherto inaccessible regimes of the two-species Hubbard model. 
First, we address the case of a small quantum bath represented by a single clean boson at weak system-bath interaction. 
Analytical arguments suggest the possibility of a breakdown of the MBL proximity effect in this regime and the delocalization of the clean particle. 
Numerically we investigate the time-evolution of the model and additionally consider a related Floquet model with similar properties, which enables the study of much longer timescales, as well as the investigation of Floquet-MBL~\cite{Abanin2015,Lazarides2015,Huse2016,Sonner2021,Sierant2023}.
At sufficiently large interaction strengths, we still observe characteristic features of localization, in contrast with the claims of Ref.~\cite{Sierant2022}.
Upon decreasing the coupling strength, we observe signatures of delocalization of the clean particle, whereas the disordered bosons still show extremely slow dynamics, making their behavior hard to capture unambiguously.

After establishing the possibility of delocalization of the single-particle bath, we investigate the effects of increasing the size of the quantum bath.
Using operator dynamics, we consider the case when the particle density of the bath and of the localized particles are comparable. 
In this case, we find non-vanishing (diffusive or weakly subdiffusive) particle transport for both, clean and disordered bosons, providing strong evidence of delocalization.
As the density of the clean particles is reduced, such that their average spacing approaches the localization length of a single clean particle in the case of the MBL proximity effect, we observe dynamics compatible with localization at short times. The rapid growth of entanglement prevents us from reaching longer timescales, where from the analytical treatment of related problems delocalization may be expected~\cite{Muller2009,Mirlin-hot-cold}. 

By pushing the limits of numerical simulations to large systems, long evolution times, and large entanglement regimes, our work sheds light on the fate of a localized system coupled to a bath with varying number of particles. 
For a weak quantum bath we demonstrate the persistence of localization in systems with a much larger number of particles than is accessible to exact diagonalization, suggesting the stability of MBL on long timescales and possibly in the thermodynamic limit~\cite{Vidmar2020,Vasseur2021b}. 
In the opposite limit of a large quantum bath, our investigation of transport shows a surprisingly fast emergence of diffusive behavior of the localized system due to its coupling with the bath, in contrast with observations of subdiffusive transport throughout the delocalized phase of more conventional disordered many-body Hamiltonians with a single species of particles~\cite{Demler2015,Znidaric2016}. 
This is suggestive of a possible effective long-range interaction induced by the clean bosons, bearing a distant analogy to studies of two-level systems coupled to waveguides that mediate long-range interactions~\cite{Chang2021}.

The paper is structured as follows.
In Section~\ref{Sec:PRACE-model} we introduce the model and give a summary of the different methods and regimes studied throughout the paper.
We then investigate the case of a quantum bath represented by a single particle, exploring the localization-delocalization transition both numerically and with an analytical approximation in Sections~\ref{Sec:PRACE-nuc0 Num} and~\ref{Sec:PRACE-nuc0 An} respectively.
Section~\ref{Sec:PRACE-nuc finite} is devoted to the case of a finite density of particles in the bath: we first study transport in the limit of large particle densities in the bath in Section~\ref{Sec:PRACE-transp}, and attempt to capture the transition as a function of particle density in the bath in Section~\ref{Sec:PRACE-nuc}.
Finally, we summarize our results and highlight possible future directions in Section~\ref{Sec:PRACE-concl}.

\section{Model}\label{Sec:PRACE-model}

We study a one-dimensional system of hard-core bosons, featuring two different particle types.
Disordered bosons, or $d$-bosons, are subject to a random on-site potential $\epsilon_i\in[-W,W]$ and are thus \textit{Anderson localized} in the absence of interactions~\cite{Anderson1958,Anderson1980}.
Clean bosons ($c$-bosons) instead have no on-site potential, with hopping amplitude $t_c$, and represent the \textit{bath}.
The two particle species are then coupled through a Hubbard type density-density local interaction, leading to the full Hamiltonian
\begin{equation}
\label{Eq:H}
\begin{split}
\hat{H} &= t_d\sum_i(\hat{d}^\dagger_{i+1}\hat{d}_i + \text{H.c.}) + \sum_i \epsilon_i \hat{n}_{d,i} \\
& + t_c \sum_i (\hat{c}^\dagger_{i+1}\hat{c}_i + \text{H.c.}) +U \sum_i \hat{n}_{c,i}\hat{n}_{d,i},
\end{split}
\end{equation}
where $\hat{n}_{c,i} = \hat{c}^\dagger_i\hat{c}_i$, and $\hat{n}_{d,i} = \hat{d}^\dagger_i\hat{d}_i$. 
The model has $U(1)\times U(1)$ symmetry due to the simultaneous particle conservation of both bosonic species.
This results in a block-diagonal structure of the Hamiltonian, with the dimension of each block determined by $ \text{dim}(\mathcal{H}) = \mathcal{C}^L_{N_d}\mathcal{C}^L_{N_c}$, with $N_{c/d} = \sum_i \hat{n}_{c/d,i}$ being the total particle number, $L$ the system size and $\mathcal{C}^m_n$ the binomial coefficient.

Throughout this paper, we study the dynamics numerically using both continuous Hamiltonian time-evolution and pulsed Floquet driving.
The Hamiltonian evolution is generated by the unitary time-evolution operator $\hat{U}(t) = \exp(-\imath \hat{H} t)$, implemented  numerically using a fourth order Suzuki-Trotter decomposition~\cite{Suzuki2005} over alternating pairs of sites with a time step $\delta t = 0.05$.
The Floquet dynamics, instead, are generated by the following time-dependent periodic Hamiltonian
\begin{equation}
\label{Eq:H(t)} 
\hat{H}(t) = \begin{cases} & \hat{H}_\text{even} \quad nT_F\leq t < (n+1/2)T_F \\
&\hat{H}_\text{odd} \quad (n+1/2)T_F\leq t< (n+1)T_F
\end{cases},
\end{equation}
where $\hat{H}_\text{even(odd)}$ represents the Hamiltonian in Eq.~(\ref{Eq:H}). Here the sum of the hopping terms is restricted to even (odd) sites, the interaction and disorder terms are halved, $T_F=0.5$ is the period and $n\in\mathbb{N}$.
The unitary Floquet operator
\begin{equation}
    \label{Eq:UF}
    \hat{U}_F = e^{-\imath \hat{H}_\text{odd}T_F/2}
    e^{-\imath \hat{H}_\text{even}T_F/2}
\end{equation}
   then describes the dynamics at \textit{stroboscopic times} $nT_F$. 
Although the Hamiltonian and Floquet models are, strictly speaking, different, the Floquet dynamics is expected to reduce to the Hamiltonian case in the limit of $T_F\to 0$. 
Moreover, both models are characterized by the \emph{exact conservation} of the particle numbers of the two bosonic species. 
This allows us to compare the transport and particle spreading between the two models.
As we demonstrate below, Floquet and Hamiltonian time evolution show similar phenomenology, with Floquet time evolution enabling us to probe the stability of MBL proximity effect and delocalization on much longer timescales at a comparable computational cost. 

\begin{figure}[t]
    \centering
    \includegraphics[width=.99\columnwidth]{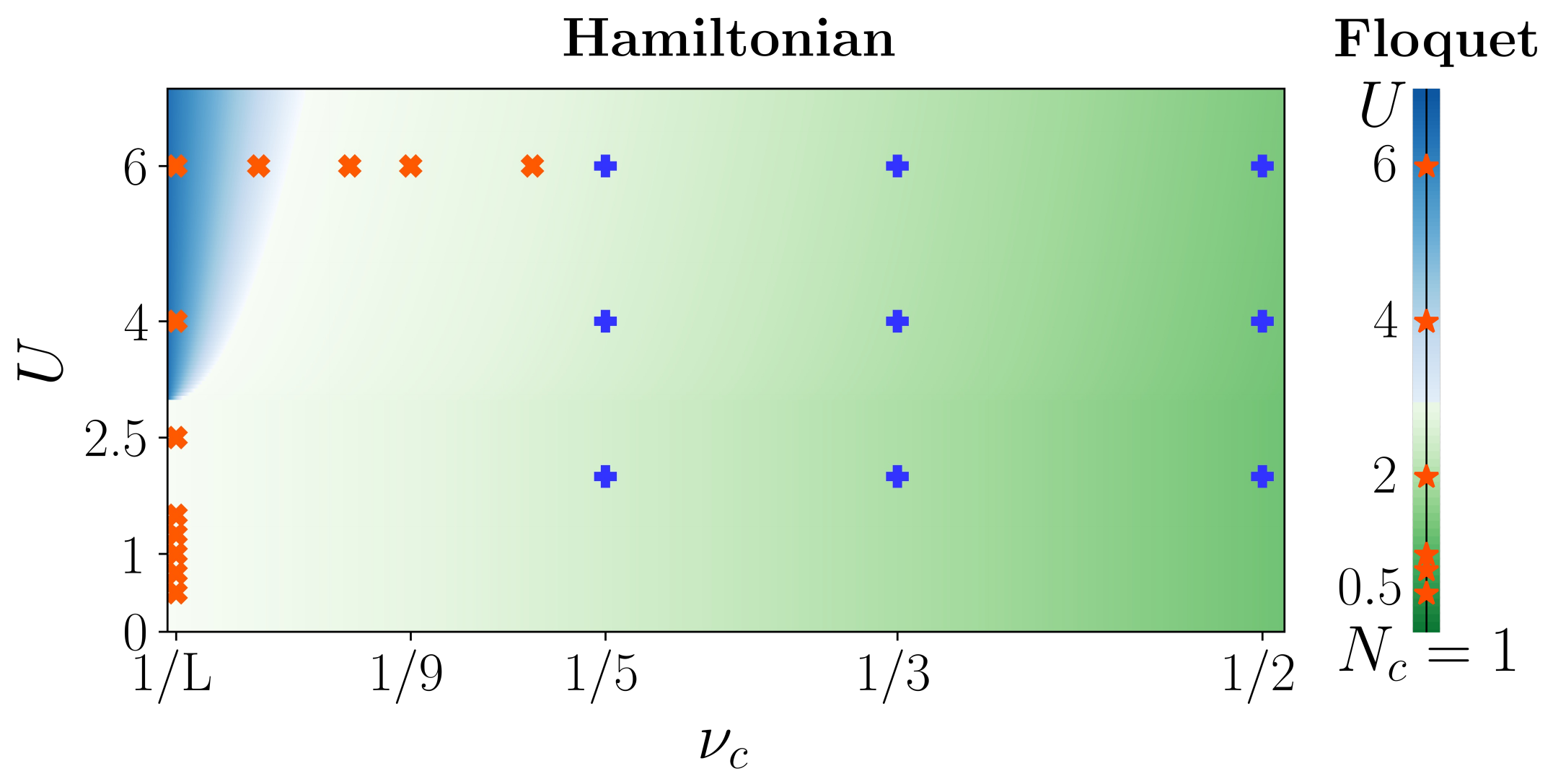}
    \caption{The phase diagram for the Hamiltonian system as function of the coupling $U$ and the bath size $\nu_c$, and for the Floquet dynamics in the case of bath consisting of a single particle ($N_c=1$ or $\nu_c=1/L$). The markers show different points in parameter space explored in this work, and refer to time evolution that uses Hamiltonian or Floquet (red markers) propagation of wave functions, and Hamiltonian evolution of density matrices (blue crosses). The colors schematically show the different phases, blue representing MBL and green the ergodic phase. The shading corresponds to the putative transition regime, obtained from our numerical simulations. Based on previous studies, we expect more stable localization in the regime of strong interactions, however in the present study we deliberately choose interactions smaller than the disorder strength, $U< W=6.5$.}
    \label{Fig:Phase Diag}
\end{figure}

To address different regimes in the parameter space of the model, we investigate the dynamics of two different types of initial conditions.
Whenever the density of clean bosons is small, $\nu_c =N_c/L\leq 1/6$, we fix the $d$-boson density to $\nu_d=1/3$ and study the evolution of the quantum wave function represented as an MPS from an initial state consisting of clean (disordered) bosons forming a density wave of period $1/\nu_{c(d)}$ respectively. 
Our numerical simulations show that upon increasing the density of clean particles in the bath the whole system approaches thermalization, thus leading to rapid entanglement spreading and making the time-evolution of the MPS wave function extremely challenging.
Thus in the regime of $\nu_c>1/6$ we time-evolve density matrices initialized close to infinite temperature ($\rho_{\infty}\propto\mathds{1}$) represented as a matrix product operator (MPO), whose simulation is more efficient in the thermal phase, see Appendix~\ref{Sec:PRACE-App extensive bath} for details.

In Figure~\ref{Fig:Phase Diag} we summarize the main results of this paper through a tentative phase diagram as a function of interaction strength $U$ and bath size controlled by the density of clean particles, $\nu_c$, for fixed parameters $t_c=t_d=1$ and $W=6.5$.
We consider interactions to be of order or smaller than disorder strength. This restriction may disfavor the localized phase that is expected to be more stable at strong interaction, but at the same time it avoids the presence of very different energy scales in the problem. 

In Section~\ref{Sec:PRACE-nuc0} we study the transition as a function of the interaction strength for the single-particle bath case, expanding our previous work~\cite{Brighi2022a,Brighi2022b} to a hitherto unexplored regime.
The joint results of our numerical simulations and analytical considerations allow us to establish the existence of two phases in this regime, an MBL phase at strong interactions (blue in the phase diagram) and a thermal phase at weak $U$ (light green), characterized by diffusive behavior of the bath particle and extremely slow relaxation of the $d$-bosons.
In this part of our study we use both Hamiltonian and Floquet dynamics, presenting qualitatively similar results.

As the number of $c$-bosons becomes extensive, $N_c\propto L$, corresponding to finite densities of clean bosons, we observe, in Section~\ref{Sec:PRACE-nuc finite}, a weakening of localization, eventually yielding delocalization shown as a crossover from blue to green in the phase diagram in Fig.~\ref{Fig:Phase Diag}. 
Finally, as $\nu_c$ is increased further we investigate particle transport deep in the delocalized region of the diagram, finding diffusive spreading of both bosonic species.

\section{Bath consisting of a single clean boson}\label{Sec:PRACE-nuc0}

We first investigate our model in the case of a bath composed of a single clean boson, $N_c=1 $.
As already shown in Ref.~\cite{Brighi2022a,Brighi2022b}, at strong interactions the disordered bosons induce localization of the small bath. 
Here we address the presence of a transition to the thermal phase as the interaction strength is decreased, using both quasi-exact large scale numerical simulations and analytical considerations.

\begin{figure*}
\includegraphics[width=0.99\textwidth]{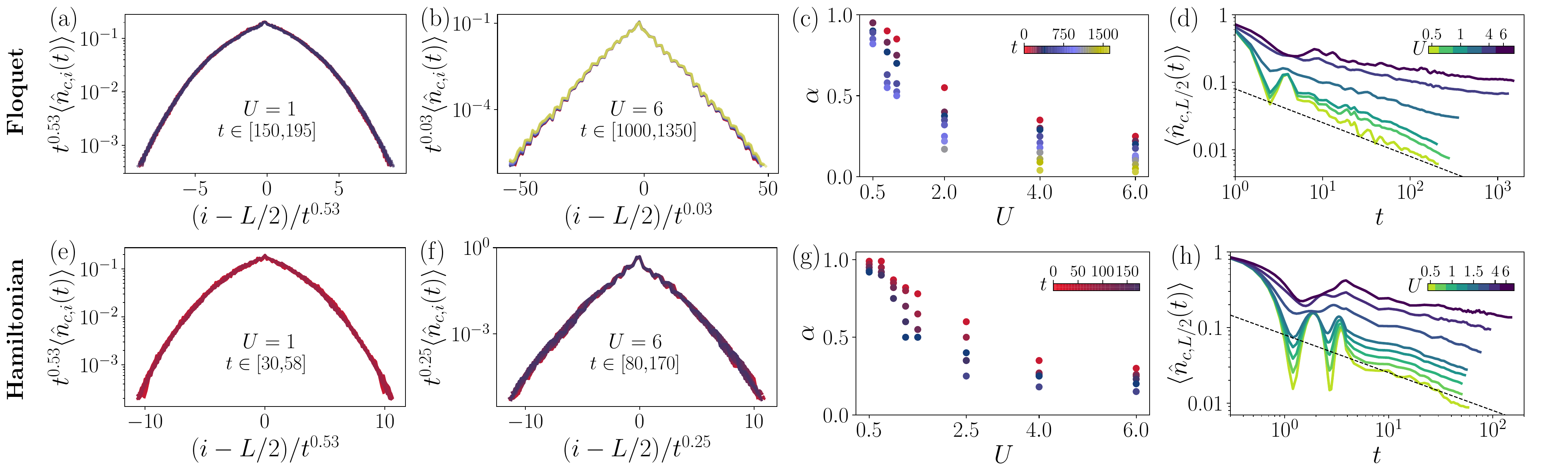}
\caption{\label{Fig:nc scaling}
In the top row we show data for the Floquet model. (a) At $U=1$ the density profiles at long times $150\leq t \leq 200$ collapse when rescaling the space axis by $\sqrt{t}$ indicating diffusive behavior of the $c$-boson.  
(b) At $U=6$, a similar collapse of the density profile is obtained with a much smaller exponent $\alpha=0.03$, a value that could be consistent with zero suggesting localization. 
(c) The exponent $\alpha(U)$ as a function of interaction strength $U$ shows that diffusive spreading of the $c$-boson at weak interactions $\alpha\approx1/2$ slows down and becomes localized at strong $U$ as is witnessed by $\alpha\to0$. The colorscale indicates the time range used to obtain the exponent, range of accessible times is limited at weak $U$ by entanglement growth. 
(d) Decay of the $c$-boson density at its original site $i=L/2$ is consistent with $1/\sqrt{t}$ (black dashed line) in the delocalized phase, and it shows signatures of saturation at strong interaction $U\geq 4$.
The bottom row shows similar data but for Hamiltonian dynamics limited to shorter times.  (e) Hamiltonian dynamics show a similar behavior indicating diffusion of the $c$-boson at weak interaction. (f) At larger $U$, density profiles collapse with a larger exponent $\alpha$ than in the Floquet case. This can be attributed to the shorter times achieved in Hamiltonian dynamics, as Floquet dynamics show comparable values of $\alpha$ at earlier times, see panels (g)-(h).
The system size is $L=2000$ for Floquet and $L=252$ for Hamiltonian dynamics, data are averaged over $10$ disorder realizations.
}
\end{figure*}

\subsection{Numerical evidence for a phase transition}\label{Sec:PRACE-nuc0 Num}

Using a highly efficient parallel implementation of the time-evolving block decimation (TEBD) algorithm~\cite{Vidal2003} with a large bond dimension $\chi=5000$ and small truncation error $\varepsilon=10^{-9}$, we simulate the dynamics generated by the Hamiltonian~(\ref{Eq:H}) in large systems of $L=252$ sites.
This choice of parameters, together with the fourth order Suzuki-Trotter decomposition with a small time-step, guarantees almost exact numerical results up to the times when bond dimension saturates.
We further explore Floquet dynamics of matrix product states of maximal bond dimension $\chi=2048$ and systems with up to $L=2000$ sites. 
The large system sizes studied and the long timescales achieved in our work allow us to exclude boundary effects due to the finite size of the system and reach the high entanglement regime. 
As we shall demonstrate below, this is particularly important  in the weak interaction case due to the delocalization and spreading of the clean boson to large distances. 
    
In both the Hamiltonian and Floquet cases, we focus on the dynamics of an initial product state corresponding to a $d$-bosons ($\blubullet$) density wave of period $1/\nu_d=3$ and a single clean boson ($\redbullet$) initialized in the middle of the system at site $i=L/2$
    \begin{equation}
        \label{Eq:psi0 nuc=1/L}
        \ket{\psi_0} = |\underbrace{\blubullet\circ\circ}_{1/\nu_d}\blubullet\circ\circ\dots\blubullet\circ\underset{L/2}{\redbullet}\blubullet\circ\circ\dots\blubullet\circ\circ\rangle.
    \end{equation}
The choice of this initial state allows us to sample the behavior of typical states where the clean boson is far from the boundaries and is surrounded by a sea of $d$-bosons of approximately uniform density.    

\begin{figure*}
    \centering
    \includegraphics[width=.99\textwidth]{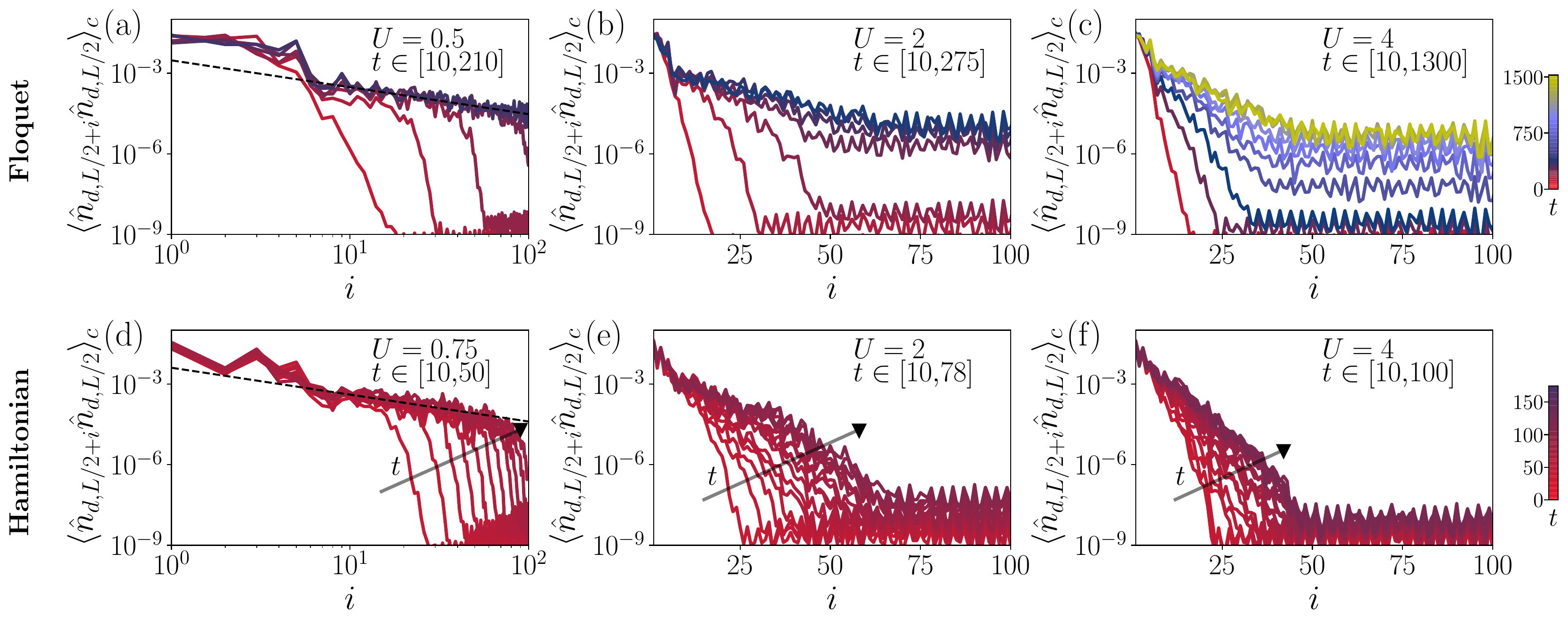}
    \caption{Density-density connected correlation functions for $d$-bosons in Floquet (a)-(c) and Hamiltonian (d)-(f) time evolution.
    At weak interactions, (a) and (d), correlations decay algebraically with distance, consistent with $1/i$ dependence (black dashed line). Such algebraic decay of connected correlation functions suggests delocalization of $d$-bosons on long timescales.
    As $U$ is increased, however, the localized behavior is eventually recovered, and panels (c) and (f) show exponentially decaying correlation functions for $U=4$.
    We notice the emergence of a plateau far from $i=L/2$ in the Floquet case, which is due to the truncation error arising from the faster saturation of bond dimension in this type of time-evolution.
    Data are averaged over $10$ disorder realizations for system size $L=2000$ and $L=252$ for the Floquet and Hamiltonian systems respectively.
    }
    \label{Fig:corrd}
\end{figure*}

\subsubsection{Diffusive behavior of the bath at weak interaction}\label{Sec:PRACE-nuc0 Num c}

We analyze the behavior of the clean particle that constitutes the bath by studying the evolution of its density profile $\langle \hat{n}_{c,i}(t)\rangle = \bra{\psi(t)}\hat{n}_{c,i}\ket{\psi(t)}$ with time.
In the localized and ergodic phases, the bath spreading is expected to show very different characteristic properties.
When the $c$-boson gets localized through the MBL proximity effect, the density profile decays exponentially away from the initial position and the localization length is expected to saturate at long times. 
In contrast, when the MBL proximity effect fails to localize the bath, the clean boson is expected to spread diffusively due to the influence of the disordered system, hence showing a Gaussian density profile with the density at the original site decaying as $\propto1/\sqrt{t}$.

To characterize the dynamics of the $c$-boson, we perform collapses of its density profile at different times, using the following scaling form:
    \begin{equation}
        \label{Eq:nc scaling}
        \langle \hat{n}_{c,i}(t)\rangle = t^{-\alpha}f\Bigr(\frac{i-L/2}{t^\alpha}\Bigr).
    \end{equation}
The value of the exponent $\alpha(U)$ can be thought as a proxy for the inverse dynamical exponent and is used to distinguish the diffusive and localized behavior of the bath.
These density profile collapses are shown in Figure~\ref{Fig:nc scaling}(a)-(b) for Floquet dynamics and in Figure~\ref{Fig:nc scaling}(e)-(f) for Hamiltonian evolution.
At weak interaction strength $U=1$, the value of $\alpha\approx0.5$ in both Hamiltonian and Floquet cases suggests a delocalized bath.
Conversely, the Floquet model at strong interaction $U=6$ exhibits a vanishing exponent $\alpha\approx 0$ highlighting the saturation of the $c$-boson spreading and suggesting the persistence of the MBL proximity effect already observed at stronger interactions~\cite{Brighi2022a,Brighi2022b}.
In the Hamiltonian time-evolution, however, the exponent $\alpha$ is still decaying and attains a larger value, that can be ascribed to the shorter times achieved in this regime.
Indeed, a comparison with the collapse in the corresponding time-window of the Floquet evolution results in a good agreement of the value of $\alpha$ as shown in Fig.~\ref{Fig:nc scaling}(c), suggesting that at later times the exponent will eventually decay also in the Hamiltonian case.

In addition to different values of $\alpha$ obtained from the rescaling, the density profiles of the clean boson in Figure~\ref{Fig:nc scaling}(a)-(b) and (e)-(f) also have a qualitatively different form. In the delocalized phase, the boson density shows a characteristic Gaussian profile, as opposed to the exponential decay observed at strong interactions.

The study of the exponent $\alpha$ at different times and as a function of $U$ is presented in Figure~\ref{Fig:nc scaling}(c) and (g).
Saturation towards $\alpha=1/2$ is observed at $U<2$, and is especially apparent for Floquet evolution. 
For Hamiltonian dynamics the values of the exponent remain close to $\alpha=1$ at times $t\approx 150$.
This suggests that a much longer time evolution is needed to see the crossover to diffusion. 
In contrast, for $U\geq2.5$ the value of $\alpha$ decreases with increasing evolution time. 
This hints at a possible transition from delocalization to localization occurs in the window of interaction strengths $1\lesssim U\lesssim 2.5$. 
Due to the fast entanglement growth observed in it, this critical region is also the most challenging to treat numerically, thus preventing a more accurate estimate of the transition point.
The transition in the behavior of the bath can also be captured by the different dynamics of the central site density decaying as $\approx 1/\sqrt{t}$ in the delocalized case and saturating to a finite value at strong $U$, as shown in Fig.~\ref{Fig:nc scaling}(d) and (h). 
Note, that the saturation of the density in the Floquet dynamics for $U=2$ in Fig.~\ref{Fig:nc scaling}(d) at long times may suggest that the bath is localized at this interaction strength, at the timescales accessible to our simulations.
Finally, in Appendix~\ref{Sec:PRACE-App nc nuc0} we provide further evidence of the $c$-boson transition by showing a diverging decay length at $U<2$.

\subsubsection{Slow delocalization of disordered bosons}\label{Sec:PRACE-nuc0 Num d}

Due to the small bath size, changing the interaction strength $U$ has a much weaker influence on the density profiles of $d$-bosons.
Indeed, comparing the density profiles and the imbalance at different values of $U$ na\"ively suggests that the disordered particles remain localized for all values of interaction strength, at least at the accessible timescales. 
A deeper investigation of more sensitive probes, provided by the entanglement entropy and the connected correlation functions of $d$-bosons, however, reveals the existence of two contrasting behaviors at weak and strong $U$. 

As shown in Appendix~\ref{Sec:PRACE-App nd+S nuc0}, the global half-chain entanglement entropy grows algebraically for  small values of the interaction, characterized by a universal behavior $S_{L/2}(t)\propto (tU^\beta)^\gamma$, with $\gamma\approx 0.39$ and $\beta\approx 1.1$ in the Floquet case.
In the Hamiltonian case a qualitatively similar picture holds, although the value of $\beta$ suddenly drops to $\approx 0.6$ as $U>1$.
This deviation from the logarithmic growth of the entanglement entropy indicates that the system cannot be fully localized.
The analysis of the entanglement profile provides further evidence in favor of delocalization, as its growth is not limited to the central part of the chain, but it propagates to regions far from the initial position of the clean particle.

To further probe the behavior of the $d$-bosons, we analyze their density-density connected correlations
\begin{equation}
        \label{Eq:corrd}
        \langle \hat{n}_{d,i}\hat{n}_{d,j}\rangle_c = \langle \hat{n}_{d,i}\hat{n}_{d,j}\rangle - \langle \hat{n}_{d,i}\rangle \langle\hat{n}_{d,j}\rangle,
\end{equation}
shown in Figure~\ref{Fig:corrd} (Floquet~(a)-(c), Hamiltonian~(d)-(f)) as a function of the distance from the center and at different times.
At weak interactions below the transition $U<U_c$, the connected correlations present a slow $1/|i-j|$ decay in space (black dashed line) and spread in time to regions far from the center, confirming the slow delocalization of the $d$-bosons.
On the other hand, in the MBL proximity effect phase, density correlations decay exponentially, with a decay length slowly increasing in time and eventually saturating, as highlighted by the collapse of the curves at late times. 
For values of $U\geq 4$ we notice that the decay length of the $d$-bosons correlations, $\ell_d(t)$, seems to saturate to a value comparable with the lengthscale of the exponential suppression of the $c$-boson density, $\ell_c(t)$.
Note that the saturation value of $\ell_{c,d} \approx 5$ obtained for $U=4$, is much smaller than the simulated system size, thus hinting at the localization of $c$ and $d$-bosons.

In conclusion, our large scale numerical simulations reveal two qualitatively different behaviors in the system at large and weak interaction strengths. Such distinction would be difficult to make in smaller systems where the finite size affects the spreading of both particle types. In the regime of large $U$, both boson species are localized. At weak $U$, while we observe a spreading of $c$-bosons over the distances of hundreds of lattice sites, the dynamics of $d$-bosons are much slower, and we are unable to fully determine their fate despite reaching long times $t\geq 200$ in our simulations.

\subsection{Estimating the critical coupling from mapping to a Bethe lattice}\label{Sec:PRACE-nuc0 An}

After identifying numerically the existence of two different phases, we construct a phenomenological picture of the transition that allows to obtain analytical estimates.
To analyze the behavior of our model in the weak interaction regime, we first consider the Hartree picture presented in Ref.~\cite{Brighi2022b}, where both the bath and the $d$-bosons are localized with localization length $\xi_c\gg\xi_d$ at weak coupling.
The Anderson orbitals $\{\ket{\alpha}\}$ form a complete basis, that we use to rewrite the particles creation and annihilation operators $\hat{d}^\dagger_i = \sum_\alpha \phi_\alpha(i)\hat{d}^\dagger_\alpha$, with $\phi_\alpha(i) = \bra{\alpha}i\rangle\approx \exp(-|i-x_\alpha|/\xi_\alpha)/\sqrt{2\xi_\alpha})$.
The interaction then corresponds to simultaneous $c$- and $d$-bosons hopping among different localized orbitals
    \begin{equation}
        \label{Eq:Hint Anderson}
        U \sum_i \hat{n}_{c,i}\hat{n}_{d,i} = \sum_{\alpha\beta\gamma\delta} V_{\alpha\beta}^{\gamma\delta}\hat{d}^\dagger_\alpha\hat{d}_\beta\hat{c}^\dagger_\gamma\hat{c}_\delta.
\end{equation}

\begin{figure}[t]
    \centering
    \includegraphics[width=.99\columnwidth]{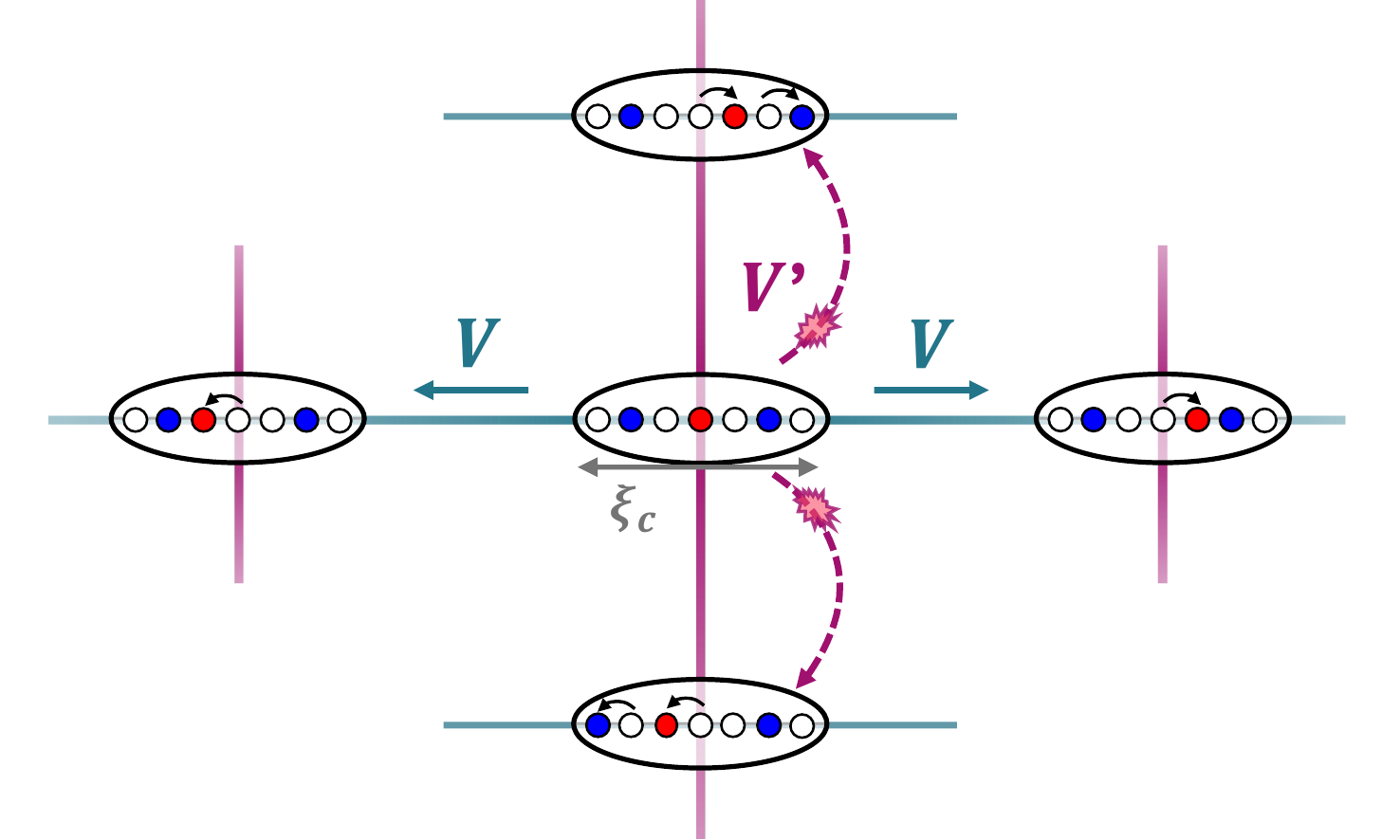}
    \caption{\label{Fig:Bethe}
    Schematic representation of the Bethe lattice, where a small $K=4$ is chosen for clarity. Each node represents a different boson configuration in Fock space. \textit{Elastic} scattering (horizontal edges) connects nodes where only the $c$-boson moves with a hopping amplitude $V$. The simultaneous hopping of clean and disordered bosons, instead, gives rise to \textit{inelastic} processes (vertical edges) with amplitude $V'$.
    }
\end{figure}

In particular, we focus on the motion of the clean boson comprising the quantum bath. 
The hopping between different orbitals, arising from the interaction term in Eq.~(\ref{Eq:Hint Anderson}) can be depicted on a graph representing the Fock space. Each node in this graph corresponds to a certain filling of localized orbitals, and edges connect different configurations with the matrix element obtained from Eq.~(\ref{Eq:Hint Anderson}). 
While long-range hoppings are allowed, they are exponentially suppressed due to the localization of both particle species in the Hartree limit.
Therefore, neglecting the edges connecting configurations where the particles move farther than their localization length, we approximate the full Fock space as a graph of connectivity $K\approx \xi_c^2\xi_d\nu_d(1-\nu_d)$, where the $\nu_d(1-\nu_d)$ term accounts for the finite density of $d$-bosons and their hard-core nature (see Ref.~\cite{Brighi2022b} for detailed derivation).

A second, crucial, approximation corresponds to neglecting all loops in the graph, resulting in a Bethe lattice with coordination number $K$, as shown pictorially in Figure~\ref{Fig:Bethe} for $K=4$. In this graph we distinguish two types of hopping processes for the clean boson, \emph{elastic} and \emph{inelastic}. 
In \emph{elastic} hoppings, the $c$-boson moves without changing the pattern of occupation of $d$-bosons orbitals, i.e. $\alpha=\beta$ in Eq.~(\ref{Eq:Hint Anderson}).
These processes, depicted by solid arrows in Figure~\ref{Fig:Bethe}, correspond to the motion of the clean-particle in the random environment created by the $d$-bosons.
Alternatively, the $c$-boson motion can simultaneously produce a scattering of $d$-boson from one orbital to another.
We refer to this second type of processes as \textit{inelastic} and represent them as dashed arrows in Figure~\ref{Fig:Bethe}.

A phenomenological mapping of the motion of the $c$-boson to a finite coordination number Bethe lattice allows us to use the results of \textcite{Abou-Chacra1973} and obtain a condition for the stability of localization.
The stability of the localized phase is controlled by the interplay of the disorder strength on the lattice $\mathcal{W}$, the connectivity, $K$, and the matrix element of the hopping processes, $V$.  Ref.~\cite{Abou-Chacra1973} derives a trascendental equation for the critical value of the matrix element $V_c$, $\frac{2KeV_c}{\mathcal{W}}\ln(\frac{\mathcal{W}}{2V_c}) = 1$, such that the localized phase is stable for $V<V_c$.
To apply this result to our model, we estimate the typical matrix element $V$, the disorder strength $\mathcal{W}$ and the connectivity $K$.
In order to establish the typical matrix element $V$, we first consider its approximate expression
\begin{equation}
    \label{Eq:V}
        V_{\alpha\beta}^{\gamma\delta} 
        \approx 
        \frac{U}{\xi_c\xi_d}\sum_i 
        e^{-\frac{|i-x_\alpha| + |i-x_\beta|}{\xi_d}}
        e^{-\frac{|i-x_\gamma| + |i-x_\delta|}{\xi_c}},
\end{equation}
where we replace the orbital-specific localization length with its average and neglect the oscillatory part of the wave function.
In Appendix~\ref{Sec:PRACE-App Bethe}, we estimate the typical value of $V$, which in the case of $\xi_c\gg \xi_d\approx 1$ can be approximated as
$V\approx {U}/({2\xi_d})$.
Finally, using the estimate for the connectivity $K\approx \xi_c^2\xi_d\nu_d(1-\nu_d)$~\cite{Brighi2022b}, we can estimate the transition by numerically solving the equation for the critical hopping amplitude on the Bethe lattice at fixed $t_c=t_d=1$ and $W=6.5$.
In the Hartree approximation the effective disorder results from the interaction with the $d$-bosons, and it is thus proportional to the coupling strength,  $\mathcal{W}\propto U$.  Also, the localization length of the $c$-boson scales as $\xi_c\propto U^{-2}$~\cite{Brighi2022b}. Thus, the decrease of the hopping amplitude at weak $U$ is counteracted by a larger effective connectivity and weaker effective disorder, leading to instability of localization below a certain critical value of interaction strength.
The numerical estimate suggests that in this parameter range, localization becomes unstable at a critical value of the coupling $U_c\approx 3$, in good agreement with the transition window inferred from the numerical results of Section~\ref{Sec:PRACE-nuc0 Num}. 

In summary, the phenomenological mapping of the hopping of the single clean boson to the Bethe lattice discussed above predicts an instability of localization at sufficiently weak interactions $U$ in agreement with our numerical simulations.
The present approach differs from the method used previously in Ref.~\cite{Brighi2022b}, where the ratio of the typical matrix element to the level spacing was used as a criterion for delocalization.
While the resulting critical curves are qualitatively similar, we expect the current mapping to the Bethe lattice to be more accurate in the weak coupling regime.
Indeed, in the present work we focus primarily on the behavior of the $c$-boson in the case of $\xi_c\gg \xi_d\approx1$, where the problem can be interpreted as a weakly localized single particle occasionally perturbed by the inelastic scattering of $d$-bosons.
Additionally, considering the motion of the $c$-boson in the Bethe lattice gives an intuitive explanation of the diffusive behavior observed in Section~\ref{Sec:PRACE-nuc0 Num}.  In the standard picture of single particle localization on the Bethe lattice, the motion of the particle in the delocalized phase is ballistic, since the majority of the steps increase the distance of the particle to the origin.  In contrast, our mapping naturally reproduces diffusion in the delocalized phase, since at each point half of the hopping processes move the $c$-boson to the left, and the remaining half moves it to the right, see Fig.~\ref{Fig:Bethe} for a schematic picture. Understanding if the present phenomenological mapping to the Bethe lattice is capable of reproducing other aspects of numerical simulations, such as entanglement dynamics or very slow relaxation of $d$-bosons in the delocalized phase remains an interesting open question. 

\section{Extensive bath}\label{Sec:PRACE-nuc finite}

In the previous Section we provided evidence of a transition between regimes of localized and delocalized small bath, tuned by the interaction strength, $U$. 
In this Section we study the effect of increasing the density of $c$-bosons that constitute the bath to a finite value. 
First, in Section~\ref{Sec:PRACE-transp} we consider the case when the density of $c$-bosons is close to half-filling, and investigate particle transport.
Afterwards, we study the regime of finite but small density of clean bosons in Section~\ref{Sec:PRACE-nuc}.

\subsection{Particle transport at large clean boson density}\label{Sec:PRACE-transp}

\begin{figure}[b]
    \centering
    \includegraphics[width=.95\columnwidth]{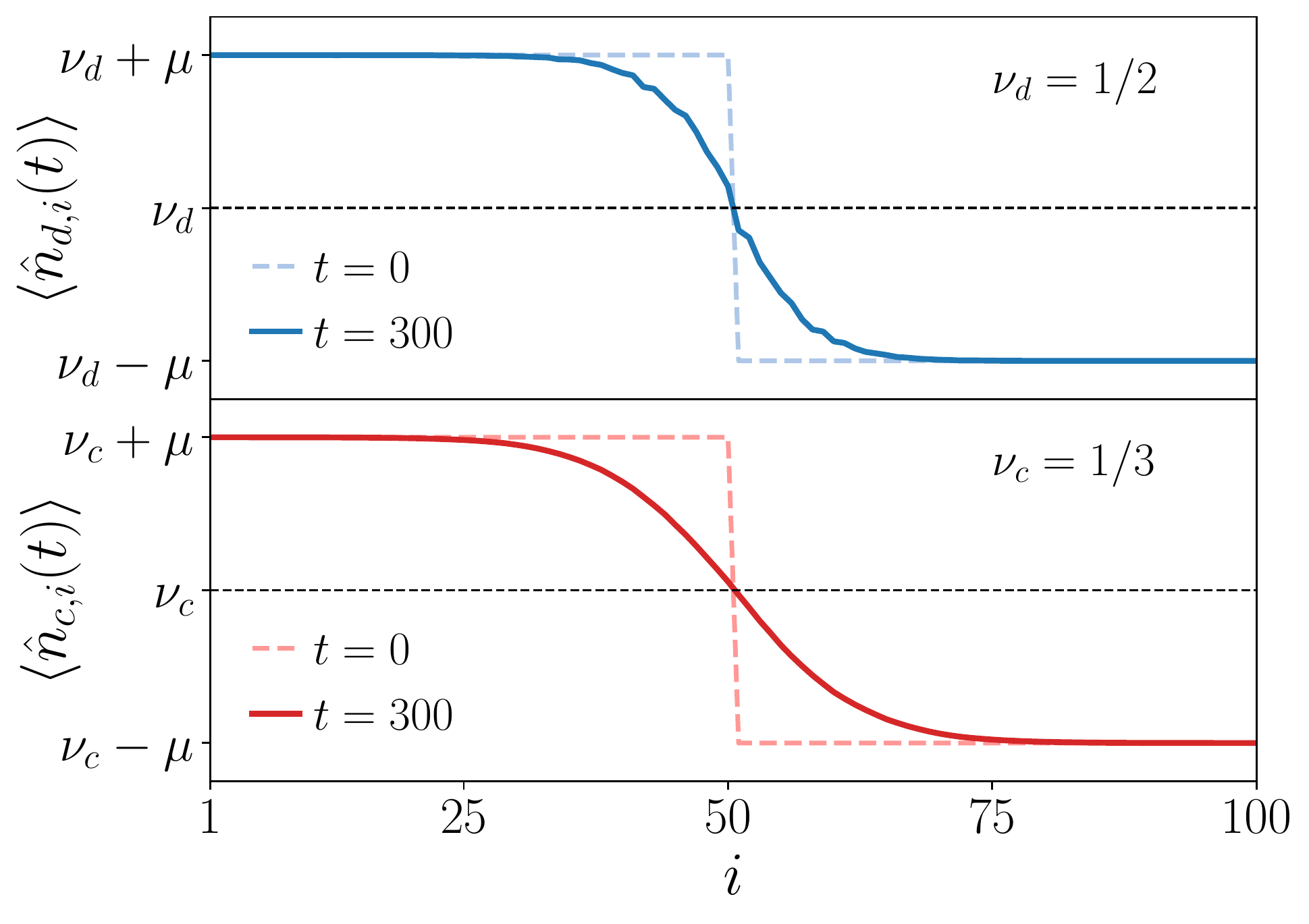}
    \caption{Density profile of disordered and clean bosons at late (solid lines) and initial (dashed lines) times. The small density step of magnitude $2\mu$ slowly melts due to the particle current running from the left to the right part of the chain. Data are shown for $U=6$, $W=6.5$ and averaged over $30$ disorder realizations.
    }
    \label{Fig:ncnd densmat}
\end{figure}

\begin{figure*}[t]
    \centering
    \includegraphics[width=.99\textwidth]{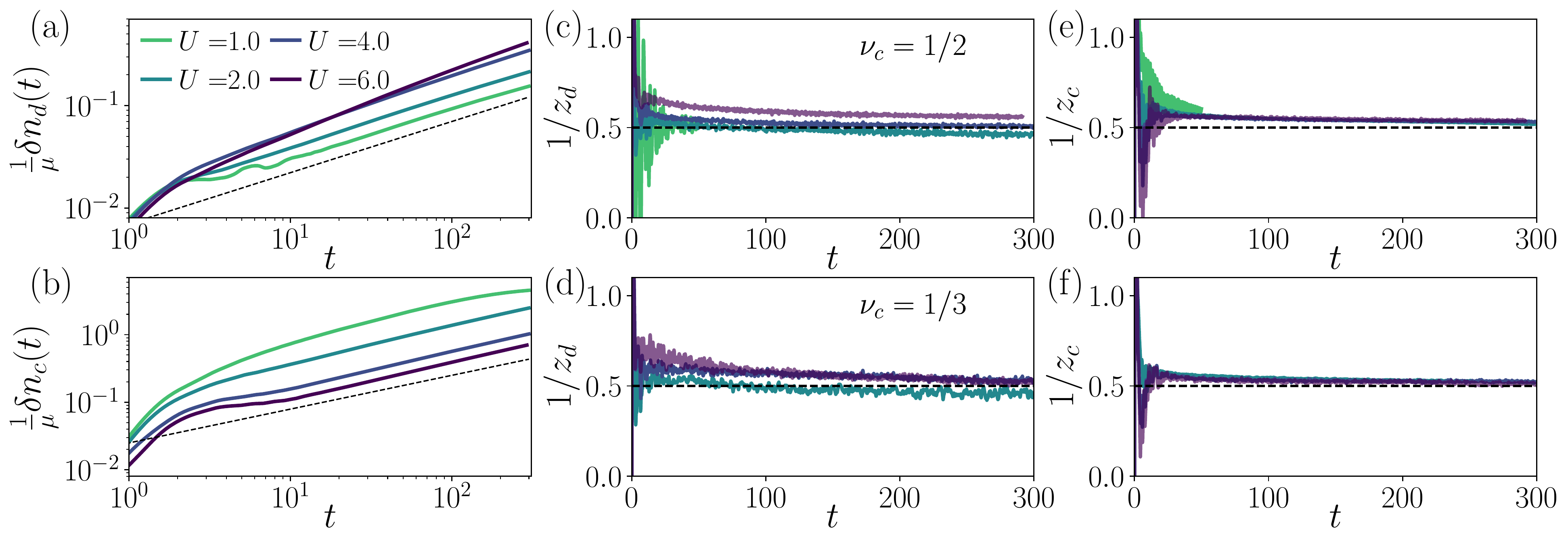}
    \caption{
    (a)-(b) Dynamics of the particle flow across the central site for $d$ and $c$-bosons at large $\nu_c=\nu_d=1/2$ and $W=6.5$. In both cases $\delta n(t)$ shows a clear power-law behavior close to diffusion $\approx \sqrt{t}$ (dashed line). However, while for the $c$-bosons $\delta n_c$ increases as the interaction strength decreases, in agreement with their free nature at $U=0$, for the disordered particles $\delta n_d$ vanishes at weak interactions approaching the Anderson localized phase.
    (c)-(f) Inverse dynamical exponent of $d$-bosons $1/z_d(t)$ and $c$-bosons $1/z_c(t)$ for different values of the bath size $\nu_c=1/2$ (c),(e) and $\nu_c=1/3$ (d)-(f). 
    At large $U$, $1/z_d(t)$ saturates to a diffusive value $z_d=2$ for both clean particle densities.
    As $U$ is decreased and the Anderson localized phase is approached, however, $d$-boson transport shows signatures of subdiffusive behavior.
    On the other hand, $c$-boson transport is almost unaffected by the coupling strength, always showing diffusive behavior, except for $U=1$.
    In this case $1/z_c$ eventually vanishes due to the finite size of the system, thus requiring larger systems to properly evaluate the dynamical exponent at late times.
    }
    \label{Fig:zdzc}
\end{figure*}

We first approach the regime where the clean particles have a large overall density $\nu_c\geq 1/5$. 
In this regime we expect that the bath triggers delocalization of the $d$-bosons. 
To characterize the resulting delocalized phase, we study the transport of bosons using the time-evolution of density matrices close to infinite temperature represented as MPOs.
This has several advantages over simulating states directly, which we discuss in Appendix~\ref{Sec:PRACE-App extensive bath}.

Following a well-established approach~\cite{Ljubotina2017}, we initialize the system in a density matrix characterized by a small step in the center of the particle density profile. Figure~\ref{Fig:ncnd densmat} illustrates such an initial density profile, with density in the left (right) part of the chain being set to $\nu_{c(d)}\pm\mu_{c(d)}$.
This condition translates to an initial density matrix written as a tensor product of density matrices on individual sites. 
The density matrix of individual site of $c$-bosons (and analogously, $d$-bosons) can be written as 
\begin{equation}
    \label{Eq:rho0}
    \begin{split}
    \rho^{(i)}_{c} = &\begin{pmatrix} 1-\nu_{c} -\mu_{c}(i)& 0 \\ 0 & \nu_{c}+\mu_{c}(i) \end{pmatrix},
    \end{split}
\end{equation}
where $\mu_c(i)=\pm0.01$ in the left and right half respectively.
Since the time-evolution of density matrices represented as MPOs is most efficient when they are close to infinite temperature, we fix the $d$-boson density to be $\nu_d=1/2$. 
We then apply operator TEBD to large chains of $L=100$ sites up to times $T=300$, using a maximal bond dimension in the range $\chi\in[128,192]$, depending on the convergence of the results, shown in Appendix~\ref{Sec:PRACE-App bond}.

To characterize particle transport, we study the evolution of the transferred particle number, $\delta n_{c(d)}(t)$, defined as the difference between the density profile at zero time and time $t$,  $\langle \hat{n}_{c(d),i}(t)\rangle$,  
\begin{equation}
\label{Eq:dn}
\delta n_{c(d)}(t) = \sum_{i=1}^{L/2}\left[\langle \hat{n}_{c(d),i}(0)\rangle - \langle \hat{n}_{c(d),i}(t)\rangle\right].
\end{equation}
The change of density with time corresponds to the current across the central site, integrated over time, thus  quantifying the transport of particles. 
In particular, the logarithmic derivative of $\delta n_{c(d)}$ with respect to time can be related to the instantaneous inverse dynamical exponent $1/z(t)$,
\begin{equation}\label{Eq:z-definition}
    \frac{1}{z(t)}=\frac{d\ln \delta n(t)}{d\ln t}.
\end{equation}

We study the particle flow as a function of interaction strength $U$ and density of particles in the bath, $\nu_c$. 
As we show in Figure~\ref{Fig:zdzc}, $\delta n_{c(d)}(t)$ at times larger than $t\geq 10$ has a clear power-law behavior for both particle species, confirming the delocalization of the originally Anderson localized $d$-bosons due to the coupling to the bath. 
However, the fact that $d$-bosons without coupling to the bath are localized, is reflected by the qualitatively different behavior of the transported number of particles with changing $U$ apparent in Fig.~\ref{Fig:zdzc}(a)-(b). 
With increasing interaction strength, the $c$-bosons exhibit slower transport, whereas the $d$-bosons are characterized by faster transport. 

In Figure~\ref{Fig:zdzc}(c)-(f), we show the instantaneous dynamical exponent $z_{c(d)}(t)$ extracted using Eq.~(\ref{Eq:z-definition}).
We report results for two different particle densities in the bath, $\nu_c=1/2$ in panels~(c)-(e) and $\nu_c=1/3$ in panels~(d)-(f), at different coupling strengths $U$.
Panels (e)-(f) show that the transport features of the bath are unaffected by the variation of the bath particle density and interaction strength, $U$, with $1/z_c(t)$ always rapidly converging to a value of $1/z_c\approx 1/2$ at late times. 
Note, that at $U=1$ and $\nu_c=1/2$ the transport of $c$-bosons is affected by the boundaries, as highlighted by the slowdown of the growth of $\delta n_c(t)$ shown in panel~(b), thus requiring larger systems to properly assess its value at late times.

In contrast to the $c$-bosons, transport of the disordered bosons is more sensitive to the choice of the parameters.
In particular, the value of the inverse dynamical exponent systematically decreases as the interaction strength and the density of particles in the bath are lowered, as indicated by the weakly subdiffusive behavior observed for $U\leq 2$.
This slight subdiffusive behavior, however, is in contrast with previous results showing strong subdiffusion in a wide parameter range in the vicinity of the transition into the ergodic phase of disordered Hamiltonians~\cite{Demler2015,Znidaric2016,Evers2017}.

In order to highlight the difference between transport observed in the present model to standard many-body localized systems, we investigate transport in the the disordered Heisenberg chain~\cite{Znidaric2008,Pal2010}.
In the two-species Hubbard model considered here interactions and hopping are of the same order, while disorder is larger and fixed to $W=6.5$. Inspired by these values of parameters, we consider the disordered Heisenberg chain with fixed disorder $W=6.5$ and hopping $J=1$ . We use interaction strength as a control parameter to tune delolcalization. Indeed, at small values of $J_z\leq1$ the model exhibits MBL properties, whereas for larger $J_z$ delocalization is expected.

Using density matrix simulations of the Heisenberg chain, as shown in Appendix~\ref{Sec:PRACE-App Heis}, we first confirm that the inverse dynamical exponent $1/z(t)$ decreases with increasing time at small values of $J_z$, consistent with localization. For larger values of $J_z$ we observe a reversal of this trend, with $1/z(t)$ increasing with time. Nevertheless, even for the longest accessible times $100\leq t\leq 250$ and for the considered broad range of $J_z=1.5\ldots 10$, the value of $1/z(t)$ remains well below one half. This is consistent with previous numerical studies~\cite{Demler2015,Znidaric2016,Evers2017} that suggested the presence of a broad subdiffusive regime even for parameter values for which the model is deep into the delocalized phase.
Since interaction, disorder strength and hopping are comparable in the two Hamiltonians, this result suggests that the presence of the bath in our model cannot simply be replaced by an effective local interaction.
Intuitively, the rapid onset of diffusion for the disordered bosons in the present model may be attributed to an effective long-range coupling among them, mediated via the particles in the quantum bath, thus providing a faster transport channel. Quantifying such emergent long-range interaction via specific observables that can be probed in TEBD time evolution remains an interesting avenue for the future work.

\subsection{Potential delocalization at small particle density in the bath}\label{Sec:PRACE-nuc}

The ergodic behavior observed in the previous Section suggests the presence of a transition as function of the bath size.
In order to capture this transition, we explore the parameter space close to the MBL phase, fixing $U=6$ and slowly increasing the $c$-boson density $\nu_c = 1/24,1/12\dots1/6$.

As the clean particle density decreases, the density matrices from Eq.~(\ref{Eq:rho0}) are too far from infinite temperature and give rise to large operator entanglement, thus rendering our method inefficient.
We hence again use MPS time-evolution, which can still capture the dynamics of the system, although the timescales may be limited by the relatively fast growth of entanglement entropy.
In this framework, we modify the initial state defined in Eq.~(\ref{Eq:psi0 nuc=1/L}) by replacing the single $c$-boson with a density wave of period $1/\nu_c$
\begin{equation}
    \label{Eq:psi0 nuc}
    \ket{\psi_0} = |\underbrace{\blubullet\redbullet\circ}_{1/\nu_d}\blubullet\circ\circ\blubullet\underbrace{\redbullet\circ\blubullet\circ\circ\blubullet}_{1/\nu_c}\redbullet\circ\blubullet\circ\circ\blubullet\redbullet\circ\blubullet\dots\rangle,
\end{equation}
so that it can accommodate an extensive number of clean bosons. 
We further choose a large system size $L=126$ such that even at the smallest density, $\nu_c = 1/24$, the bath hosts a significant number of clean bosons,  $N_c=5$.

In Appendix~\ref{Sec:PRACE-App finite nuc}, we report on the behavior of density profiles and imbalance, confirming delocalization at large $\nu_c$ characterized by relaxation of the initial density wave pattern.
At smaller bath densities, however, signatures of thermalization are absent up to timescales $T\approx50$.
In order to capture the transition, then, we analyze the behavior of a more sensitive probe -- entanglement entropy.

\begin{figure}[t]
    \centering
    \includegraphics[width=.99\columnwidth]{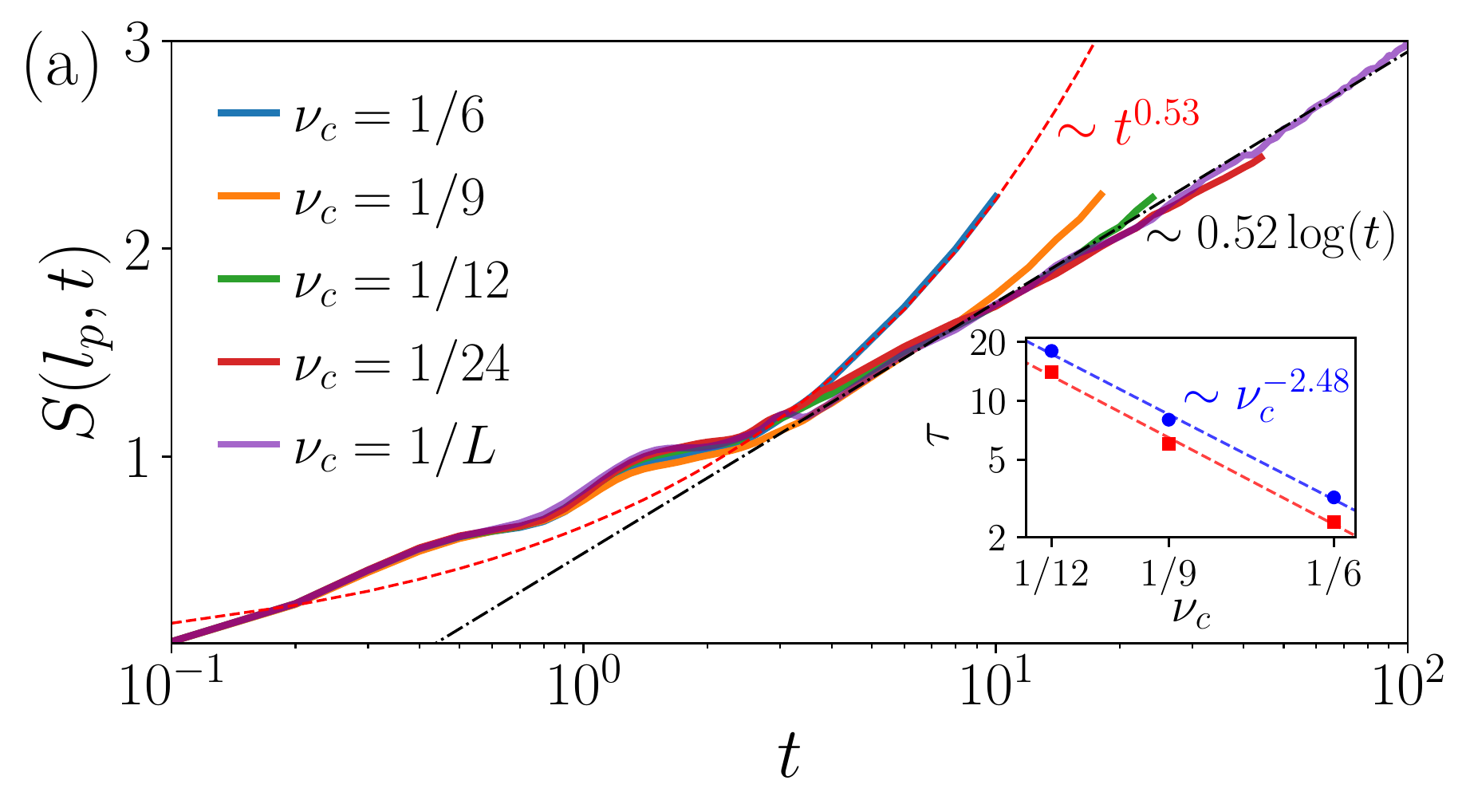}\\
    \includegraphics[width=.99\columnwidth]{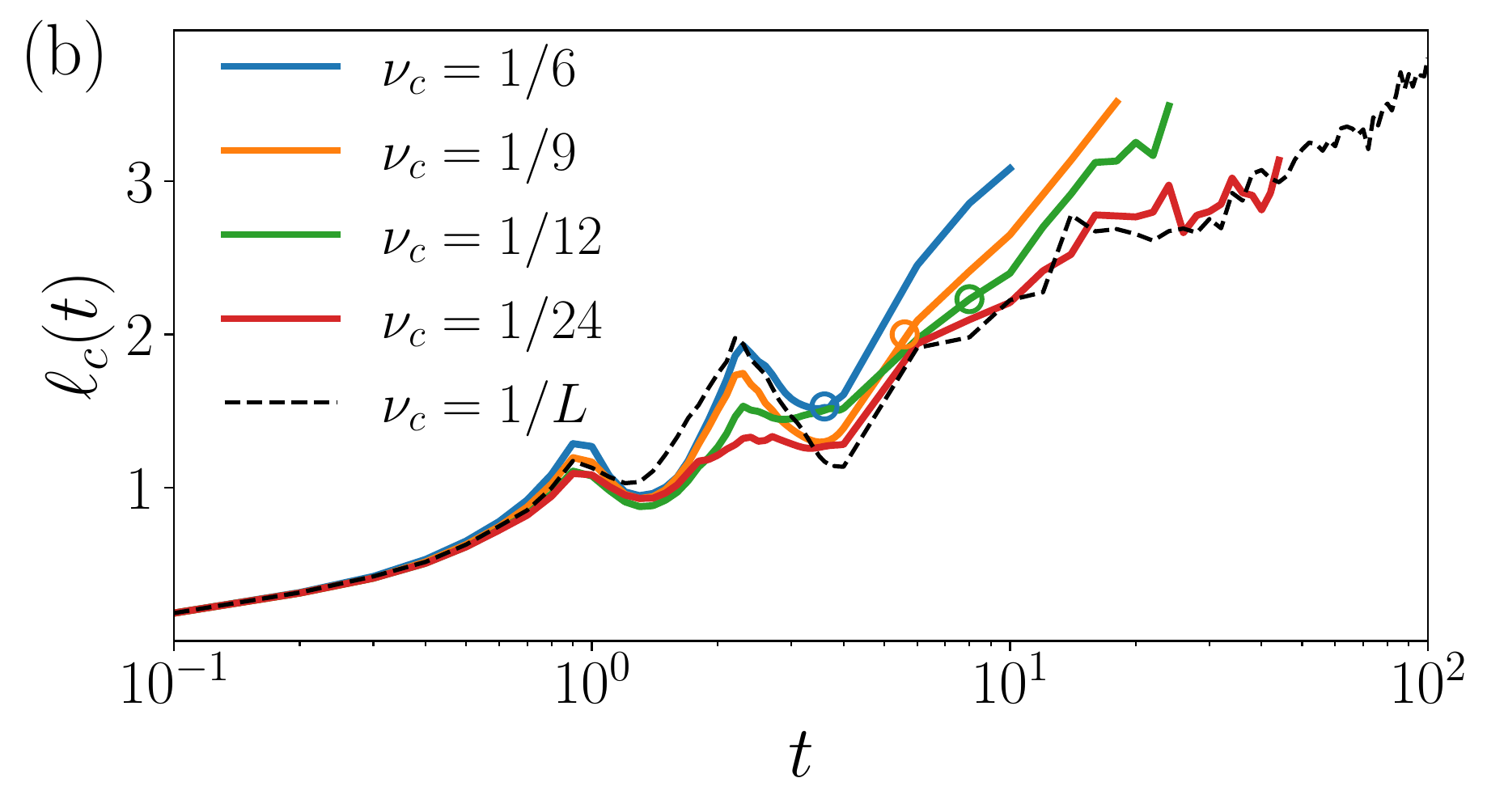}
    \caption{(a): Entanglement entropy across a cut between sites $l_p$ and $l_p+1$, where $\langle \hat{n}_{c,l_p}(t=0)\rangle=1$ for all the densities studied.
   For  densities $\nu_c\geq1/12$ we observe a deviation at a time $\tau_S(\nu_c)$ from the curve corresponding to the case of a single $c$-boson (purple line).
    In the inset we show the behavior of this characteristic timescale as a function of density, suggesting a power-law behavior.
    Comparison of $\tau$ obtained from the entanglement entropy (blue dots) and the one obtained from the growth of the correlation functions (red squares) suggests that both timescales behave in a similar way.
    (b): The decay length $\ell_c(t)$ averaged among all $c$-bosons composing the bath also deviates from the decay length of a single clean boson (black dashed line) at late times for  $\nu_c\geq1/12$.
    }
    \label{Fig:S nuc}
\end{figure}

Since entanglement growth is influenced by the distance to the closest $c$-boson, we compare bipartite entanglement across a cut $l_p$ such that $\langle \hat{n}_{c,l_p}(t=0)\rangle=1$ for all considered densities of clean bosons.
This comparison is shown in Figure~\ref{Fig:S nuc}(a), together with the entanglement entropy relative to the bath consisting of a single clean particle.
After an initial transient logarithmic regime, entanglement entropy eventually curves upwards, indicating a faster power-law growth with time.
We observe such deviation from the single-particle case for all values $\nu_c\geq 1/12$ and track its characteristic onset time $\tau_S(\nu_c)$.
The scaling of $\tau_S(\nu_c)$ is shown in the inset of Fig.~\ref{Fig:S nuc}(a). Within the considered range of densities, the behavior of $\tau_S(\nu_c)$ is consistent with a power-law increase at low $\nu_c$, $\tau_S(\nu_c)\propto 1/\nu_c^{-2.48}$, that implies eventual delocalization at any finite density of clean bosons. Of course, the limited data range does not allow us to rule out the possibility that $\tau_S(\nu_c)$ diverges at a finite value of $\nu_c$, signaling stability of localization at a finite density of clean bosons.  

The analysis of the density-density correlation functions of the $c$-bosons shown in Appendix~\ref{Sec:PRACE-App finite nuc}, provides a possible explanation for the power-law entanglement growth. 
We notice that the correlations among two initially occupied sites $l_p$ and $l^\prime_p$ become relevant at a timescale scaling with the bath density with the same power-law exponent observed for $\tau_S$, as shown by the red line in the inset of Fig.~\ref{Fig:S nuc}(a).
The agreement between the scaling of the two timescales implies that the faster entanglement growth is triggered only when $c$-boson correlations become significant. As a consequence of the strong suppression of the spreading of the clean bosons degrees of freedom, when $\nu_c$ is small enough correlations may saturate to a very small value compatible with the logarithmic growth of entanglement, thus possibly leading to a localized phase even at a finite density of clean bosons. 

Deviations from single-particle behavior can also be observed in the time-evolution of the $c$-boson density profiles.
In particular, we probe this by studying the decay length $\ell_c(t)$ of each individual $c$-boson obtained by fitting the density profile in the vicinity of each $l_p$ to the ansatz proposed in Ref.~\cite{Brighi2022a}
\begin{equation}
\label{Eq:PRACE nc(x,t)}
n_{c}(x,t) = \mathcal{N}_c(t)\exp\Bigr(-\frac{|x|}{\ell_c(t)\tanh\bigr(\frac{R(t)}{|x|}\bigr)}\Bigr).
\end{equation}
Here $\mathcal{N}_c(t)$ is a time-dependent normalization factor and $R(t)$ describes the Gaussian part of the profile emerging at small densities.
In the present case, we average the results for $\ell_c(t)$ obtained for each clean boson.
As shown in Figure~\ref{Fig:S nuc}(b), at early times the decay length behaves in agreement with the single-particle case for all densities $\nu_c$ (note that the deviations from the dashed curve can be attributed to the additional averaging for finite $\nu_c$).
At later times marked by circles, however, the decay length grows consistently faster than in the case of an intensive bath for all bath sizes $\nu_c\geq 1/12$.

While the scaling of $\tau$ shown in the inset of Fig.~\ref{Fig:S nuc}(a) seems to suggest thermalization at all finite $\nu_c$, the potential localization of the individual $c$-bosons indicated by the agreement of the decay length for $\nu_c=1/24$ with the single $c$-boson curve might lead to a breakdown of the delocalization mechanism at small, albeit extensive, bath sizes.
Unfortunately, on the timescales available to our numerical simulations, we are not able to confirm that the $\nu_c=1/24$ bath leads to power-law growth of entanglement or to deviations in the growth of $\ell_c(t)$.
Using the power-law fit for $\tau_S(\nu_c)$, we can estimate that the deviation from logarithmic growth at small bath size would take place at a time $T\approx 100$, corresponding to a nearly uniform  entanglement entropy $S\geq 3$ -- an extremely challenging regime in TEBD simulations.

\section{Discussion}\label{Sec:PRACE-concl}

Our work highlights novel physical aspects of localization that can be studied using multi-species lattice models. While considering the mixture of two hard core bosons severely impacts the system sizes reachable with exact diagonalization techniques, it presents a smaller limiting factor for the highly efficient MPS-based numerical simulations employed here. In the present case, considering the two species model resulting in a mixture of disordered and clean particles allowed us to show the stability of the MBL proximity effect~\cite{Nandkishore2015a}.
Besides the persistence of localization at strong coupling, we also presented evidence of a transition driven by the interaction strength. Decreasing the interaction leads to delocalization and diffusion of the clean particle, together with a slow inhomogeneous relaxation of the disordered bosons.

In addition, we explored the presence of a phase transition tuned by the density of clean bosons. 
Our study of transport highlights that at large densities of clean particles the system is delocalized.
The investigation of the putative delocalization transition at small but finite density of clean bosons, however, is not conclusive due to the rapid entanglement growth preventing our simulation from reaching sufficiently late times. Thus, further studies are needed in order to understand if there exists a finite critical density of clean particles below which the entire system stays localized~\cite{Nandkishore2015a}, or whether delocalization occurs at any finite density of clean particles akin to scenario suggested by Refs.~\cite{Muller2009,Mirlin-hot-cold}.

Numerous other questions remain open. In particular, although the single clean particle spreads diffusively at weak interactions, the disordered bosons show extremely slow relaxation. Such behavior is intuitively similar to the delocalized yet non-ergodic phase suggested to exist on Bethe lattices~\cite{Ioffe2018,Khaymovich2022,Mirlin2021}. It remains to be understood, if our phenomenological mapping of the two-species Hubbard model to the Bethe lattice can reproduce the relaxation of disordered bosons, entanglement dynamics, and other physical properties, such as the behavior of connected correlation functions. More broadly, the delocalization of a single particle atop of the infinite sea of localized bosons presents a deviation from the standard thermodynamic limit, where typically densities of all particle species are assumed to be finite. Building a theory of the delocalization transition for such a system remains an interesting challenge. Furthermore, the role of interactions between disordered particles as well as the nature of disorder (random or quasiperiodic) provide a complementary set of control parameters, that was not explored in this work. 

In a different direction, the consideration of finite density of clean particles allows for a standard thermodynamic limit, but turns out to be an extremely challenging problem for numerical simulations. In this regime, although we were able to confirm delocalizaiton at large densities of clean particles, the intermediate density regime proved to be hard to access due to rapid entanglement growth. Understanding the structure of this entanglement and searching for a quasi-local basis transformation may potentially assist one in reaching longer simulation times. This may be crucial for getting insights into microscopic processes and the structure of resonances created by the clean particles that drive the delocalization of the entire system. Additionally, further investigation of the diffusive transport at large densities of particles in the bath may provide useful insights for a better understanding of the subdiffusion observed in standard disordered models used to study many-body localization~\cite{Demler2015,Znidaric2016,Evers2017}.  

Finally, modern experiments with ultracold atoms motivate the study of other geometries and setups for the many-body lozalization proximity effect. In particular, the study involving the two dimensional cold atom microscope~\cite{Rubio-Abadal2019}, that inspired our work, calls for an extension of our results to two-dimensional systems. While simulating dynamics with two-dimensional tensor network ansatzes is extremely challenging, the study of the present model on ladders with MPS methods may provide useful insights into the qualitative difference between one- and two-dimensional systems~\cite{Mirlin2020} Likewise, large scale numerical studies of models where the coupling between the localized system and the bath is local~\cite{Leonard2020} may provide useful insights for the theory of many-body localization and its potential instabilities known as bubbles or avalanches~\cite{DeRoeck2017,DeRoeck2018,Luitz2017,Huse2022}. 
    
\begin{acknowledgments}
We thank A.~A.~Michailidis and A.~Mirlin for insightful discussions. 
P.~B., M.~L., and M.~S. acknowledge support by the European Research Council (ERC) under the European Union's Horizon 2020 research and innovation program (Grant Agreement No.~850899).
D.~A. was supported by the European Research Council (ERC) under the European Union’s Horizon 2020 research and innovation programme (grant agreement No.~864597) and by the Swiss National Science Foundation.
P.B., M.L. and M.S. acknowledge PRACE for awarding us access to Joliot-Curie at GENCI@CEA, France, where the TEBD simulations were performed. 
The TEBD simulations were performed using the ITensor 
library~\cite{itensor}. 
\end{acknowledgments}

\appendix

\section{Intensive bath}\label{Sec:PRACE-App nuc0}

In the main text, we showed numerical results suggesting the existence of a transition to a delocalized phase as the interaction strength between the bath and the disordered particles is decreased.
In this Appendix, we provide some additional numerical results for the dynamics of the bath, of the $d$-bosons and of the system as a whole.

\begin{figure}[t]
\includegraphics[width=.99\columnwidth]{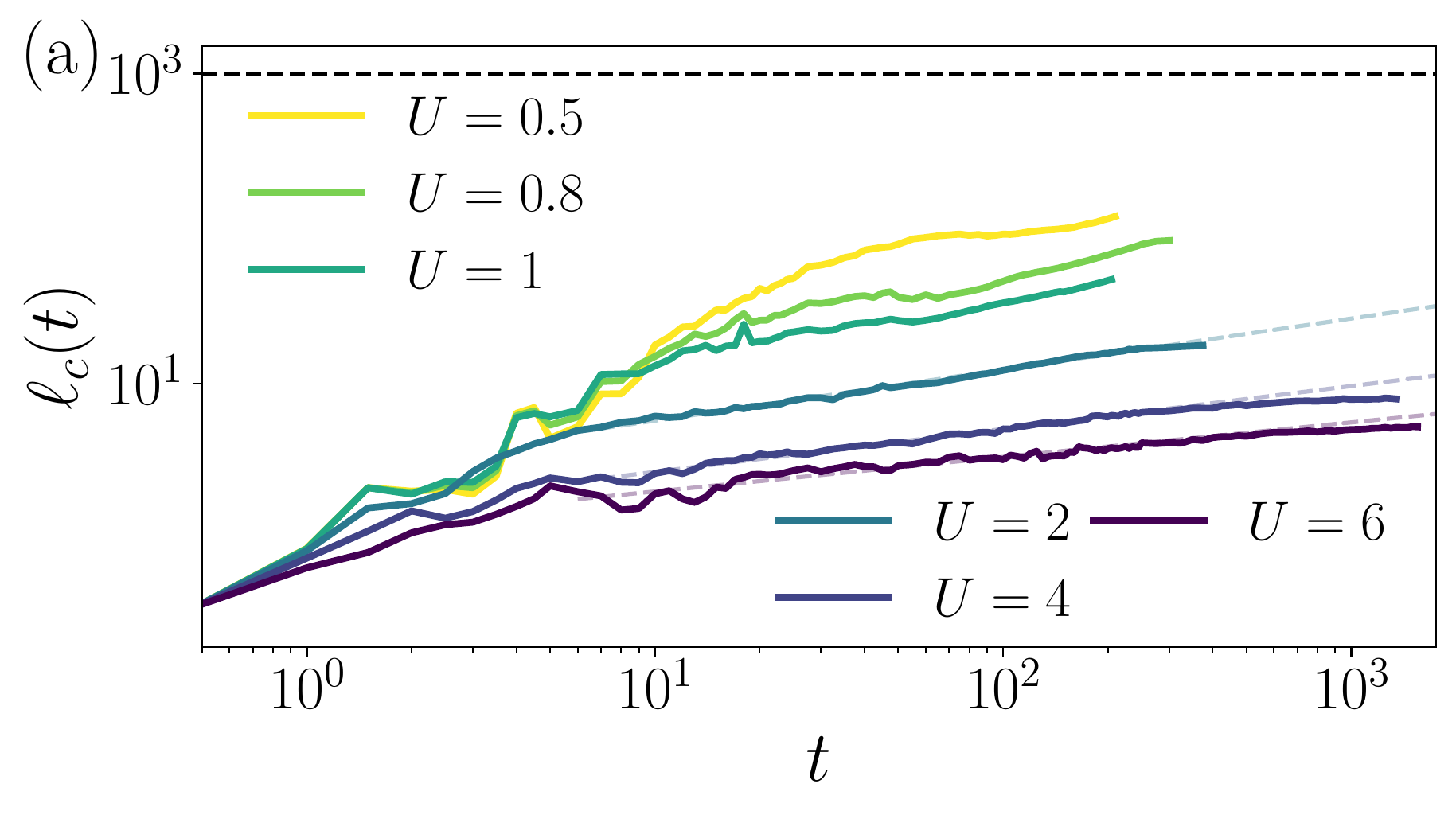}\\
\includegraphics[width=.99\columnwidth]{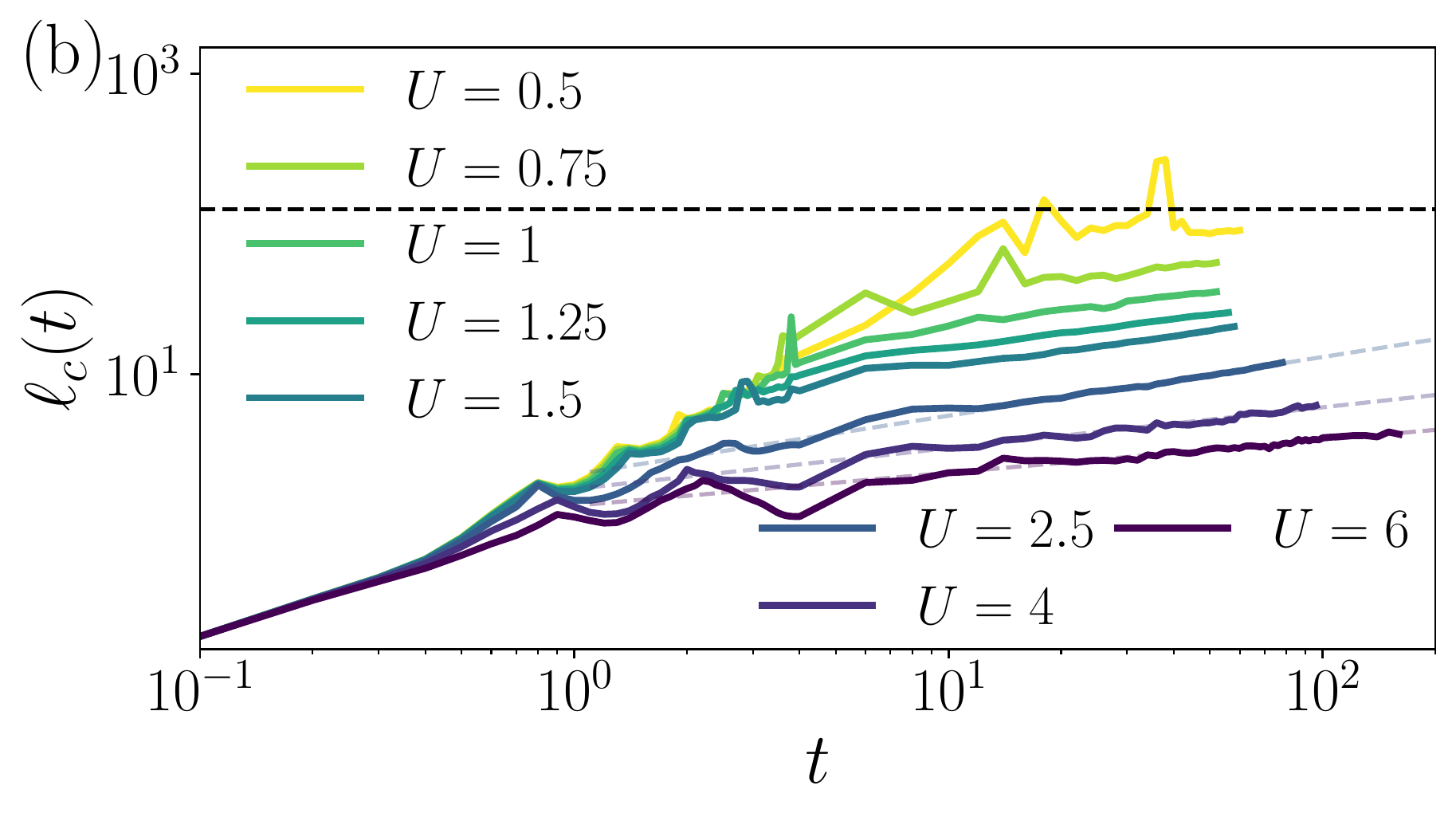}
\caption{\label{Fig:PRACE-App ellc}
Clean boson decay length $\ell_c(t)$ shows different behavior at weak and strong interactions both in the Floquet~(a) and Hamiltonian~(b) time evolution.
At small values of $U$, below the estimated transition, $\ell_c(t)$ grows persistently with no signs of saturation,
reaching the value of half system size $L/2=126$ for the Hamiltonian dynamics (black dashed line).
As the interaction is increased, the decay length first grows as a power-law in time (shaded dashed lines), but eventually, at  the long timescales accessible in Floquet evolution, starts saturating. This saturation suggests the potential stability of localization at large $U$.
}
\end{figure}

\subsection{Evaluation of the accuracy of TEBD simulations}\label{Sec:PRACE-App bond nuc0}

Due to the finite bond dimension $\chi$ and the finite truncation error $\varepsilon$, the wavefunction evolved using the TEBD algorithm $\ket{\psi_\chi(t)}$ deviates from the true state $\ket{\psi(t)}$.
The error produced by the truncation of the singular values $\lambda_a$ and corresponding Schmidt states, however, is well controlled and can be easily evaluated.
At each singular value decomposition, the local weight of the wavefunction lost is given by 
\begin{equation}
    \label{Eq:PRACE-App local error}
    \epsilon_n = \sum_{\lambda_a<\varepsilon} \lambda^2_a \quad \text{or}\quad \epsilon_n = \sum_{a>\chi} \lambda^2_a
\end{equation}
depending on whether the bond dimension at the evaluated link is saturated or not.

The global error at each timestep, then, corresponds to
\begin{equation}
    \label{Eq:PRACE-App: global err}
    \epsilon(t) = 1-|\bra{\psi(t)}\ket{\psi_\chi(t)}|^2 = 1-\prod_{n}(1-\epsilon_n).
\end{equation}
The total discarded weight at the end of the time evolution finally amounts to the integrated error $\epsilon(t)$.

In the work presented in the main text, we use extremely large bond dimensions and low truncation error, thus ensuring the accuracy of our simulations.
In the Hamiltonian case we set $\chi=5000$ and $\varepsilon=10^{-9}$, resulting in a maximum truncation error of $\epsilon = 2\times 10^{-5}$ arising close to the putative transition, at $U=1.25$.
In the Floquet simulations we use a smaller bond dimension $\chi=2048$ and the same truncation as for the Hamiltonian evolution.
This gives rise to a maximum error of $\epsilon=8\times 10^{-3}$, again at interaction strength $U=1$ in the vicinity of the transition.

\subsection{Additional evidence of bath delocalization}\label{Sec:PRACE-App nc nuc0}

In Section~\ref{Sec:PRACE-nuc0 Num}, we studied the density profile of the clean boson, highlighting the stark difference at weak and strong interactions.
We now use the ansatz for the density introduced in Ref.~\cite{Brighi2022a}
\begin{equation}
\label{Eq:PRACE-App nc(x,t)}
n_{c}(x,t) = \mathcal{N}_c(t)\exp\Bigr(-\frac{|x|}{\ell_c(t)\tanh\bigr(\frac{R(t)}{|x|}\bigr)}\Bigr)
\end{equation}
to study the behavior of the decay length $\ell_c(t)$ as the interaction is changed.

\begin{figure*}
    \centering
    \includegraphics[width=.95\textwidth]{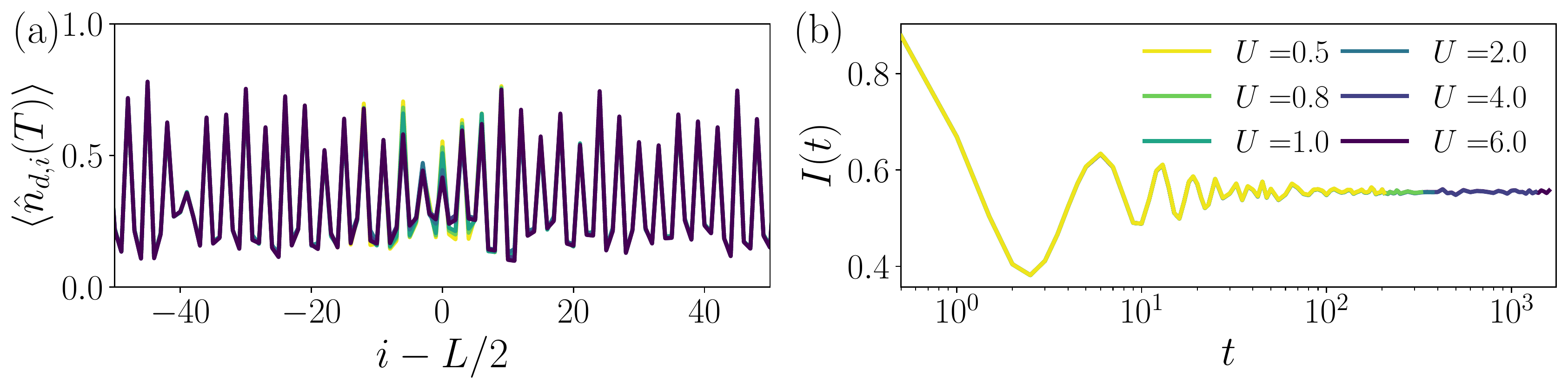}\\
    \includegraphics[width=.95\textwidth]{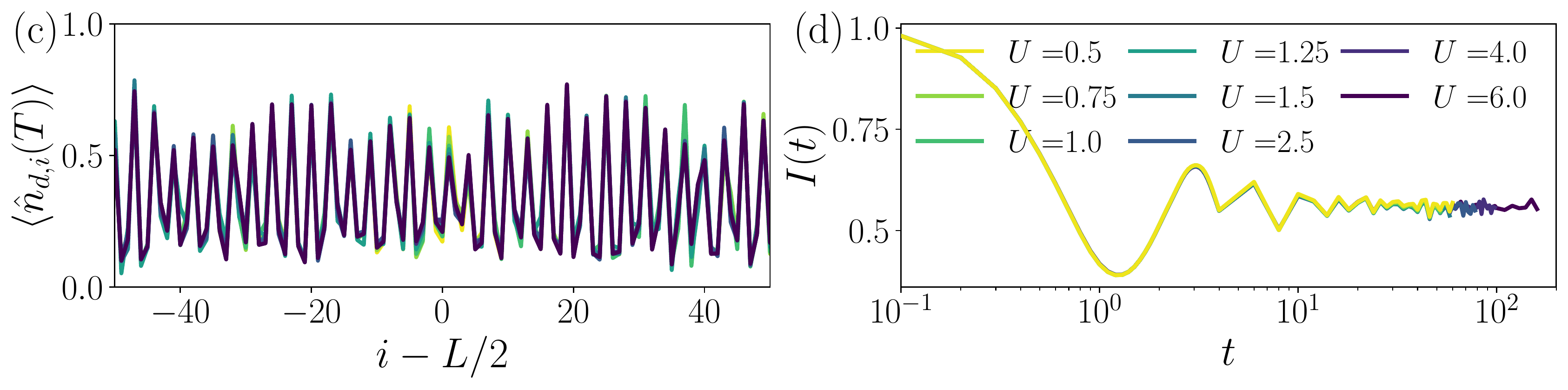}
    \caption{
    Late time $d$-boson density profiles [shown for the latest common time among different values of $U$, $T=205$ and $T=54$ for Floquet~(a) and Hamiltonian~(c) dynamics respectively] show very weak dependence on the interaction strength. The slight   enhancement of relaxation at the center of the chain at large $U$ is caused by the increased effect of localized $c$-bosons on disordered particles. 
    The dynamics of imbalance $I(t)$ in panels (b) [Floquet] and (d) [Hamiltonian] also illustrates nearly complete absence of density pattern relaxation.
    }
    \label{Fig:PRACE-App nd}
\end{figure*}

In Figure~\ref{Fig:PRACE-App ellc}, the dynamics of the decay length is presented for both the Floquet~(a) and the Hamiltonian~(b) models.
The data show a continuous increase of the value of $\ell_c(t)$ as interaction is decreased, in agreement with the transition predicted.
In particular, the decay length shows signatures of saturation to a value much smaller than the system size for $U\geq2$ and $U\geq4$ in the Floquet and Hamiltonian dynamics respectively.
At weaker interactions, instead, it keeps growing, which is suggestive of delocalization. 

\subsection{Disordered boson imbalance and global entanglement}\label{Sec:PRACE-App nd+S nuc0}

Due to the vanishing density of the bath particle in the extremely large systems we study, its effect on the $d$-boson density is weak and hard to capture from the study of their density alone.
However, in Section~\ref{Sec:PRACE-nuc0 Num} we demonstrate that a study of the density-density connected correlations allows one to observe a qualitative change in the behavior of the disordered particles at weak interactions.

We now show numerical results concerning the density profiles and the relative imbalance, to highlight the necessity of studying more complicated operators, such as the connected correlations, and entanglement entropy, to distinguish the behavior of the system at weak and strong interactions.
In Figure~\ref{Fig:PRACE-App nd}~(a)-(c) we show the $d$-boson density profiles in Floquet (Hamiltonian) dynamics at late times $T=205$ ($T=54$), respectively.
A monotonously enhanced relaxation is observed close to the central site $i=L/2$  as the interaction increases. 
This observation that the density of localized bosons is more affected by the clean particle at strong interactions (when the clean particle is localized) is readily explained by the fact that the confinement of the clean boson to a smaller region helps to relax the density pattern of disordered particles within the localization volume. Far from the center of the chain, however, no significant variation is observed as the interaction strength is changed.

A more quantitative understanding can be obtained from the study of the imbalance~\cite{Bloch2017a,Bloch2017b}
\begin{equation}
\label{Eq:PRACE-App I}
I(t) = \frac{N_o(t)-N_e(t)}{N_o(t)+N_e(t)},
\end{equation}
where $\hat{N}_{o/e}$  represent the density in the initially occupied/empty sites.
For a period $1/\nu_d=3$ density wave, they read
\begin{equation}
\label{Eq:PRACE-App No/e}
\hat{N}_o = \sum_{i=1}^{L/3} \hat{n}_{d,3i-2},\;\; \hat{N}_e = \frac{1}{2}\sum_{i=1}^{L/3}(\hat{n}_{d,3i}+\hat{n}_{d,3i-1}).
\end{equation}
As imbalance measures the \textit{memory} of the initial condition, its vanishing implies delocalization.
However, as can be seen in panels~(b)-(d) of Figure~\ref{Fig:PRACE-App nd}, both in Floquet and Hamiltonian dynamics the imbalance does not show any sign of decay on the accessible timescales, irrespective of the interaction strength.
This suggests that the relaxation of the $d$-bosons quantified by their density pattern is extremely slow.

\begin{figure*}
    \centering
    \includegraphics[width=.95\textwidth]{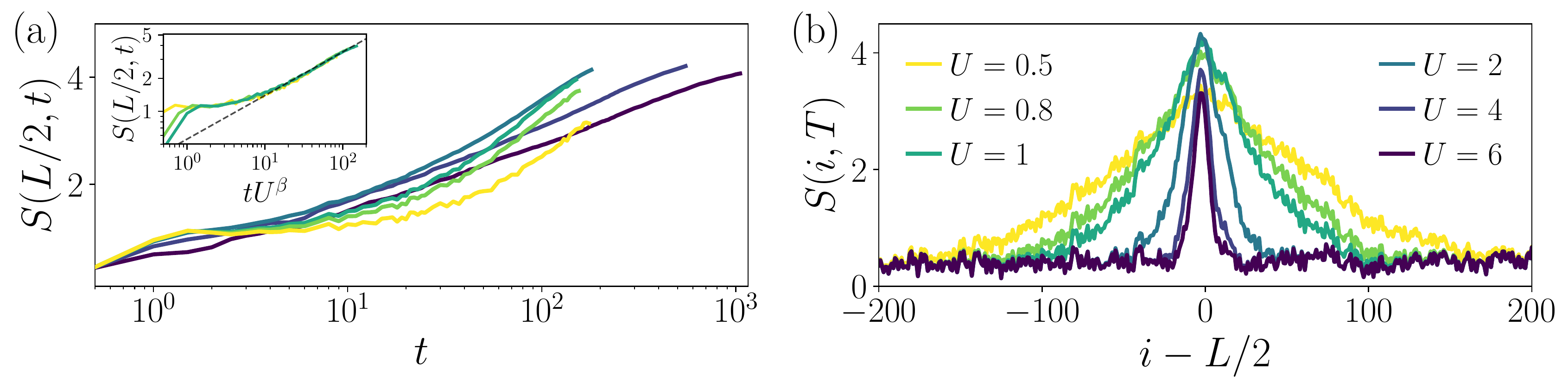}\\
    \includegraphics[width=.95\textwidth]{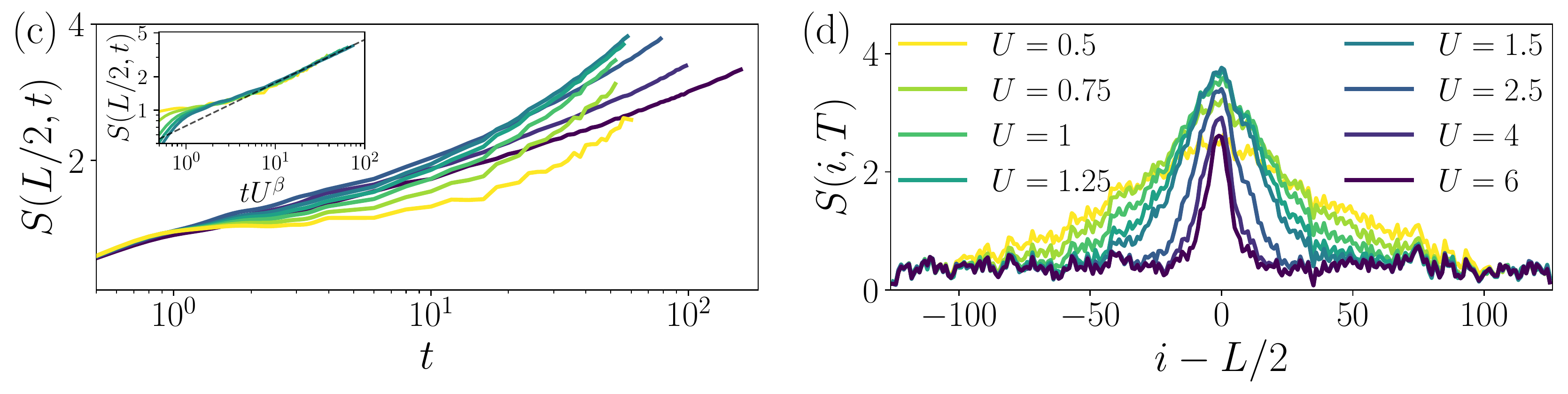}
    \caption{Dynamics of entanglement entropy show signatures of delocalization for weak coupling $U$ both in Floquet~(a)-(b) and Hamiltonian~(c)-(d) dynamics. 
    The power-law growth of entanglement across the central cut shown in panels~(a)-(c) at weak coupling  is clearly distinct from the logarithmic behavior observed for $U\geq 4$.
    The different behavior is also highlighted by the entanglement collapse shown in the inset, suggesting a universal power-law behavior in the ergodic phase.
     The behavior of real-space entanglement profiles at a fixed time [$T=205$ for Floquet~(b) and $T=52$ for Hamiltonian dynamics~(d)] in panels (b)-(d) also shows non-monotonic behavior, with a peak in the region $U\in[1,2]$ separating two distinct regimes.
     At weaker interactions entanglement spreads more uniformly through the chain, suggesting the presence of a large ergodic region close to the center.
    In contrast, at strong coupling, entanglement growth remains limited to the center of the chain, in agreement with the phenomenology of the MBL proximity effect~\cite{Brighi2022a}.
    \label{Fig:PRACE-App S}
    }
\end{figure*}

The study of the entanglement entropy 
\begin{equation}
\label{Eq:PRACE-APP S}
S(i,t) = -\tr \rho_A(t)\ln \rho_A(t),
\end{equation}
where the chain is split into two subsystems $A=[1,i]$, $B=[i+1,L]$, and we consider density matrix $\rho_A(t) = \tr_B\ket{\psi(t)}\bra{\psi(t)}$,  provides further evidence in favor of the existence of a transition.
In many-body localized systems, entanglement entropy is expected to grow logarithmically~\cite{Bardarson2012,Serbyn2013a}, hence a deviation from the logarithmic behavior can be interpreted as a breakdown of MBL.
As we show in Figure~\ref{Fig:PRACE-App S}~(a)-(c) entanglement entropy at weak interactions shows faster than logarithmic behavior, suggesting thermalization.
In particular, this leads to a universal power-law scaling of entanglement given by $S\approx (tU^\beta)^\gamma$ with the value of $\gamma\approx 0.39$ for both Floquet~(a) and Hamiltonian~(b) case.
The value of $\beta$ is similar between the two models $\beta\approx1.1$, however only for $U\leq 1$. For larger values of $U$ up to $1.5$, $\beta$ stays the same in the Floquet case and abruptly changes to $\beta \approx 0.6$ (not shown) in the Hamiltonian case.
The universal behavior is captured by the collapse of the different entanglement curves shown in the inset of Fig.~\ref{Fig:PRACE-App S}~(a)-(c).
As the interaction strength increases logarithmic growth is eventually restored at $U=4,6$, in agreement with MBL.

The different behavior is also highlighted by the spatial profile of entanglement, i.e. the entanglement entropy shown as a function of the size of subsystem $A$ at a fixed time $T$ ($T=205$ in panel~(b) and $T=52$ in panel~(d)). 
Figure~\ref{Fig:PRACE-App S}~(b)-(d) shows that at strong interactions entanglement growth is limited to the center of the chain, in agreement with the phenomenology proposed in Ref.~\cite{Brighi2022a}.
At weaker interactions entanglement starts spreading more uniformly through the chain, indicating the creation of large ergodic regions in the chain.
Interestingly, this phenomenon gives rise to a non-monotonicity of the peak of the entanglement profile as a function of interaction strength.
At very weak interactions, entanglement spreads to far regions, and in turn its value in the center is lower than at intermediate $U$, where its spreading is reduced.
We observe a maximum of the peak around $U\approx 2 $ in the Floquet case and $U\approx 1.5$ in the Hamiltonian dynamics, which could correspond to the transition point, separating the two entanglement regimes.
The value $U$ at which we observe the largest entanglement appears to be stable with respect to the fixed time $T$ for all accessible times $T>10$.

\subsection{Hopping matrix element for effective mapping to Bethe lattice}\label{Sec:PRACE-App Bethe}

In the main part of the text, we used the approach of Ref.~\cite{Abou-Chacra1973} to estimate the critical interaction strength separating the localized and ergodic phases.
In this appendix, we detail the evaluation of the typical matrix element, crucial in estimating the transition.
As mentioned in the text, the functional form of the matrix element Eq.~(\ref{Eq:V}) depends on the relative position of the different particles involved in the process.
In particular, they can be either \textit{mixed}, i.e. $dcdc,\; dccd,\;cdcd,\;cddc$, or they can be \textit{ordered}, corresponding to $ddcc,\;ccdd$.

Let us first consider the mixed case, and analyze the case $cdcd$ in detail.
We define the different positions of the particles as $x_{c,d}^{(1,2)}$ where the numerical indices represent the ordering from left to right.
It will be useful to define the distance among same type particles $r_{c(d)}=x_{c(d)}^{(2)}-x_{c(d)}^{(1)}$ and the centre of mass $\overline{x}_{c(d)}=(x_{c(d)}^{(2)}+x_{c(d)}^{(1)})/2$.
In this setup, we identify three different regions: region $(I)$: $j\leq x_d^{(1)}$, region $(II)$: $x_d^{(1)}<j<x_c^{(2)}$, and region $(III)$: $j\geq x_c^{(2)}$.
In region $(I)$ the summand is $f(j)\leq e^{-r_c/\xi_c}e^{-2(\overline{x}_d-j)/\xi_d}$, in region $(II)$ $f(j)$ is constant and attains its maximal value $f(j) = e^{-rc/\xi_c-r_d/\xi_d}$, finally in region $(III)$ $f(j)\leq e^{-r_d/\xi_d}e^{-2(j-\overline{x}_c)/\xi_c}$.
Consequently, the sum in Eq.~(\ref{Eq:V}) can be split into three terms
\begin{equation}
    \label{Eq:V cdcd}
    \begin{split}
        V &\leq \frac{U}{\xi_c\xi_d} \bigr[e^{-r_c/\xi_c}\sum_{j\leq x_d^{(1)}}e^{-2(\overline{x}_d-j)/\xi_d}\\
        &+ e^{-rc/\xi_c-r_d/\xi_d}(d_{cd}-2) + e^{-r_d/\xi_d}\sum_{j\geq x_c^{(2)}}e^{-2(j-\overline{x}_c)/\xi_c}\bigr],
    \end{split}
\end{equation}
where we introduce $d_{cd}=x_c^{(2)}-x_d^{(1)}$ as the distance between the two middle particles.

The two sums now can be rewritten, introducing a geometric series
\begin{equation}
    \label{Eq:Sum cdcd}
    \begin{split}
        &\sum_{j\leq x_d^{(1)}}e^{-2(\overline{x}_d-j)/\xi_d} = \sum_{j\leq x_d^{(1)}}e^{-(r_d-2(j-x_d^{(1)})/\xi_d} \\
        &= e^{-r_d/\xi_d}\sum_{k\geq0} e^{-2k/\xi_d} = \frac{e^{-r_d/\xi_d}}{1-e^{-2/\xi_d}}.
    \end{split}
\end{equation}
A similar approach can be used for the second sum in Eq.~(\ref{Eq:V cdcd}).
The matrix element for the case of $cdcd$ configuration can thus be estimated to be
\begin{equation}
    \label{Eq:V cdcd final}
    V\approx \frac{U}{\xi_c\xi_d}e^{-r_d/\xi_d-r_c/\xi_c}\bigr[\frac{1}{1-e^{-2/\xi_d}}+(d_{cd}-2) + \frac{1}{1-e^{-2/\xi_c}} \bigr].
\end{equation}
With a similar approach, we can also obtain the upper bound for the matrix element in the remaining cases of mixed bosons, leading to the same result for the $dcdc$ case and to 
\begin{equation}
    \label{Eq:V cddc final}
    V \approx \frac{U}{\xi_c\xi_d}e^{-r_d/\xi_d-r_c/\xi_c}\bigr[\frac{2}{1-e^{-2/\xi_{c(d)}}}+(d_{cc(dd)}-2) \bigr]
\end{equation}
for the $dccd$ ($cddc$) cases.
We numerically check that, in most cases, the bound is reasonably tight, yielding a very small percentage difference from the actual value.

\begin{figure}[b]
\includegraphics[width=.99\columnwidth]{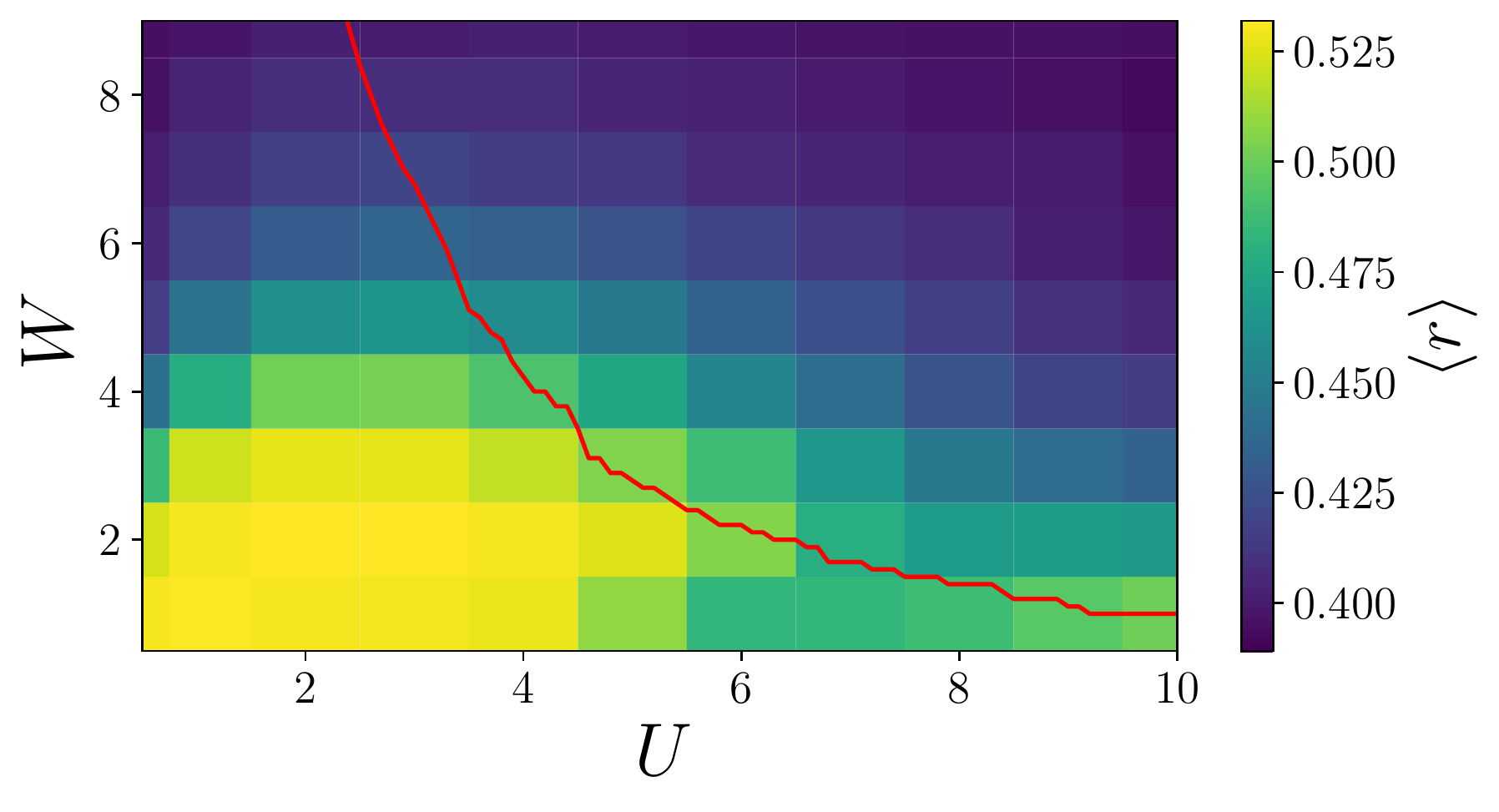}
\caption{\label{Fig:PRACE-App Wc}The critical curve obtained from solving Eq.~(\ref{Eq:PRACE-App crit}) numerically (red line) compared with the level spacing ratio for a small $L=15$ system.
Our analytical approach predicts a transition at a critical interaction strength $U_c\approx 3$ for the disorder $W=6.5$ used in the numerical simulations.
This estimate is compatible with the transition window suggested by the results of our large scale numerical simulations.
}
\end{figure}

We now consider the case of ordered particles.
In this case, the summand $f(j)$ is peaked at the position of the middle boson with shorter localization length.
Again, for the sake of clarity, we illustrate a particular configuration, but an analogous approach can be used for the other configurations.
We choose to show the results for the case $ddcc$ where the localization length of $d$-bosons is shorter than the one of the $c$-boson (the most typical case).
In this setup we can identify two different regions: region $(I)$, $j\leq x_d^{(2)}$, where $f(j)\leq e^{-r_d/\xi_d}e^{-2(\overline{x}_c-j)/\xi_c}$, and region $(II)$, $j>x_d^{(2)}$, where $f(j)\leq e^{-2(j-\overline{x}_d)/\xi_d-2(\overline{x}_c-j)/\xi_c}$.
The matrix element in region $(I)$ is upper bounded by
\begin{equation}
    \label{Eq:V ddcc (I)}
    \begin{split}
        V^{(I)} &\leq \frac{U}{\xi_c\xi_d}e^{-r_d/\xi_d}\sum_{j\leq x_d^{(2)}}e^{\frac{2j-x_c^{(1)}-x_c^{(2)}}{\xi_c}} \\
        &= \frac{U}{\xi_c\xi_d}e^{-r_d/\xi_d-r_c/\xi_c-2d_{dc}/\xi_c}\sum_{j\leq x_d^{(2)}}e^{\frac{2(j-x_d^{(2)})}{\xi_c}} \\
        &= \frac{U}{\xi_c\xi_d}\frac{e^{-r_d/\xi_d-r_c/\xi_c-2d_{dc}/\xi_c}}{1-e^{-2/\xi_c}}.
    \end{split}
\end{equation}
In region $(II)$ we obtain
\begin{equation}
    \label{Eq:V ddcc (II)}
    \begin{split}
        V^{(II)}&\leq \frac{U}{\xi_c\xi_d}\sum_{j\geq x_d^{(1)}+1} e^{\frac{-2j+x_d^{(1)}+x_d^{(2)}}{\xi_d}}e^{\frac{2j-x_c^{(1)}-x_c^{(2)}}{\xi_c}} \\
        &= \frac{U}{\xi_c\xi_d}e^{-\frac{r_d+2}{\xi_d}}e^{-\frac{r_c+2d_{dc}-2}{\xi_c}}\sum_{k\geq0}e^{-2\frac{\xi_c-\xi_d}{\xi_c\xi_d}k} \\
        &= \frac{U}{\xi_c\xi_d}\frac{e^{-\frac{r_d+2}{\xi_d}}e^{-\frac{r_c+2d_{dc}-2}{\xi_c}}}{1-e^{-2\frac{\xi_c-\xi_d}{\xi_c\xi_d}}}.
    \end{split}
\end{equation}
The case for $\xi_c<\xi_d$ can be obtained in the same way, bearing in mind that now $f(j)$ is peaked at $j=x_c^{(1)}$, and yields the same results as Eqs.~(\ref{Eq:V ddcc (I)})-(\ref{Eq:V ddcc (II)}), except with subscripts for $c$- and $d$-bosons exchanged.
A similar statement is valid for the case of $ccdd$ ordered bosons.

\begin{figure*}
    \centering
    \includegraphics[width=.95\textwidth]{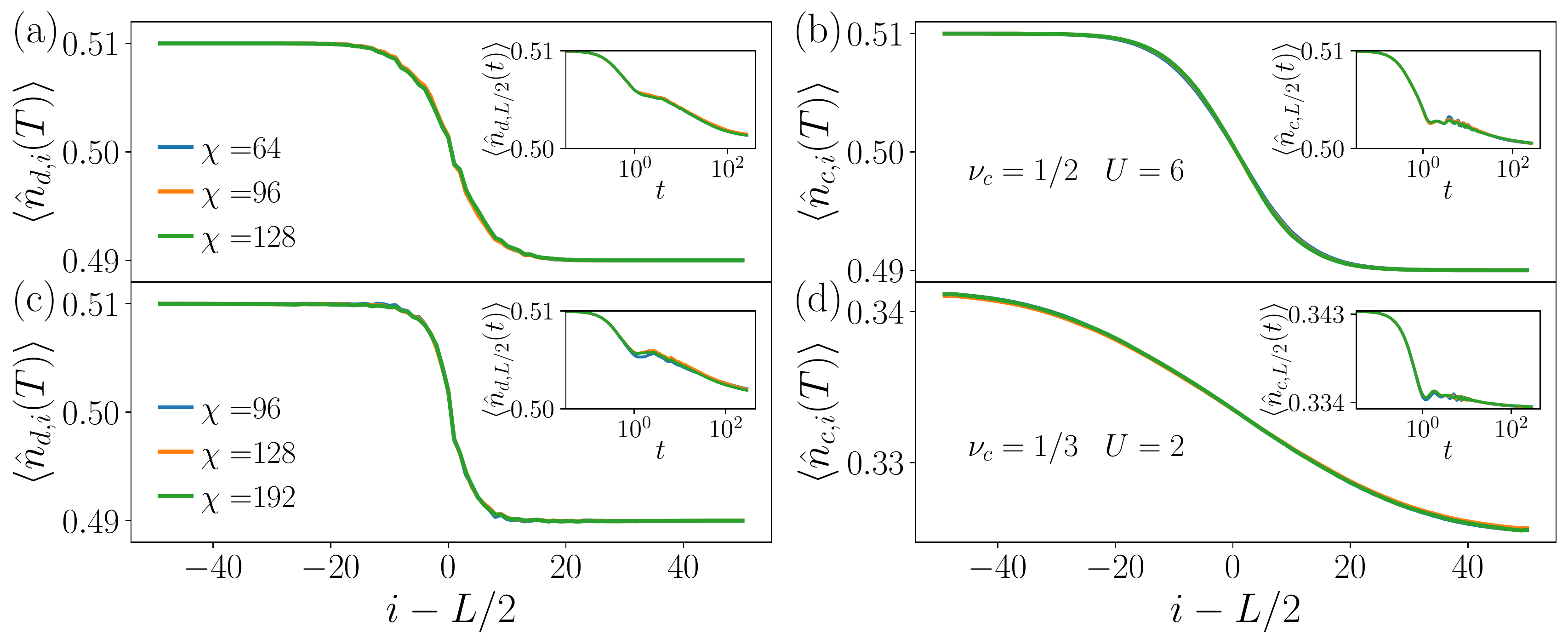}
    \caption{
    Bond dimension comparison for two exemplary values of the interaction strength and density, $U=6$, $\nu_c=1/2$ in panels~(a)-(b) and $U=2$, $\nu_c=1/3$ in panels~(c)-(d).
    The main panels show the density profiles for $d$- and $c$-bosons at the latest time $T=300$, confirming the convergence of our results for the bond dimensions used.
    Additionally, in the insets we compare the time-evolution of the density at the central site, which also shows good convergence with increasing bond dimension.
    }
    \label{Fig:PRACE-App bond}
\end{figure*}

Now the estimate of the typical matrix element corresponds to evaluating the average distances $r_{c(d)}$ and $d_{cd}$, which results in $\langle r_{c(d)}\rangle \approx \xi_{c(d)}/2$
and $d_{cd}\approx \min(\xi_c,\xi_d)$.
Finally, using the localization length obtained numerically in the Hartree approximation~\cite{Brighi2022b} and solving the equation for the critical point
\begin{equation}
\label{Eq:PRACE-App crit}
\frac{2KeV_c}{\mathcal{W}}\ln\Bigr(\frac{\mathcal{W}}{2V_c}\Bigr)=1
\end{equation}
numerically, we draw the critical line shown in red in Figure~\ref{Fig:PRACE-App Wc}.
In the heat map we additionally show results for the average level spacing ratio $\langle r \rangle$~\cite{Oganesyan2007,Pal2010} obtained from exact diagonalization of the Hamiltonian~(\ref{Eq:H}) on $L=15$ sites.
From this approximate analysis we extract a critical value of the coupling $U_c\approx 3$ for the disorder strength $W=6.5$ used throughout this work.
This result is somewhat counterintuitive, given that the hopping amplitude is proportional to the interaction, $V\propto U$. Nevertheless, as already obtained in a previous work~\cite{Brighi2022b}, the effective disorder for the $c$-boson is proportional to the coupling strength, and hence its localization length $\xi_c\propto U^{-2}$. 
This results in the observed trend yielding delocalization at weak $U$ and localization at strong interactions.

\section{Extensive bath}\label{Sec:PRACE-App extensive bath}

When studying the behavior of the model with a finite density of $c$-bosons we employ two distinct approaches. 
At relatively low densities we use MPS simulations with the $c$-bosons initially equidistant to one another.
However at higher densities this approach will lead to a prohibitively fast growth of entanglement entropy, which slows down numerical simulations and prohibits us from reaching long timescales.
To avoid this issue, we employ an MPO-based approach at large densities around half-filling, where we simulate the dynamics of the system's density matrix.
In particular, we prepare an initial density matrix with a small step in the $c$- and $d$-boson densities, which can be written as a trivial MPO with bond dimension one.
Provided that the local boson density remains near half-filling (such as $1/2$ or $1/3$) this initial density matrix leads to relatively slow growth of operator space entanglement entropy (as was also observed in an earlier work~\cite{Ljubotina2017}).
This allows us to reach the timescales necessary to extract information on the transport properties of our model at a high density of $c$-bosons. 

\subsection{Bond dimension comparison for density matrix TEBD}\label{Sec:PRACE-App bond}

In order to estimate the accuracy of our simulations, we compare the value of certain observables throughout time-evolution for different bond dimensions.
In Figure~\ref{Fig:PRACE-App bond} we present the different density profiles of the $d$-bosons and the bath for different values of the interaction and different sizes of the bath.
In the top row we show $U=6$ and $\nu_c=1/2$, while panels~(c)-(d) present $U=2$ and $\nu_c=1/3$.
The density profiles at $T=300$, shown in the main panels, are converged at the bond dimensions used in the simulations presented in the main text.
Additionally, in the insets we plot the time-evolution of the density at the central site for different bond dimensions, similarly showing converged dynamics at the largest bond dimensions.

\subsection{Transport in the disordered Heisenberg chain}\label{Sec:PRACE-App Heis}

The clear diffusive behavior observed in the transport of $d$-bosons in Section~\ref{Sec:PRACE-transp} is not commonly observed in disordered systems.
Previous studies~\cite{Demler2015,Znidaric2016,Evers2017} have reported slow, subdiffusive, transport in the ergodic phase of disordered Hamiltonians.

In this Appendix, we explore dynamics in the disordered Heisenberg chain in a parameter range where disorder and hopping are comparable to the ones investigated in our work.
The aim of this study is to check whether there exists a coupling strength such that the nearest-neighbor interaction of the disordered Heisenberg chain can reproduce diffusive dynamics in a similar timescale as the one observed in the main text.

\begin{figure}[b]
    \centering    \includegraphics[width=.99\columnwidth]{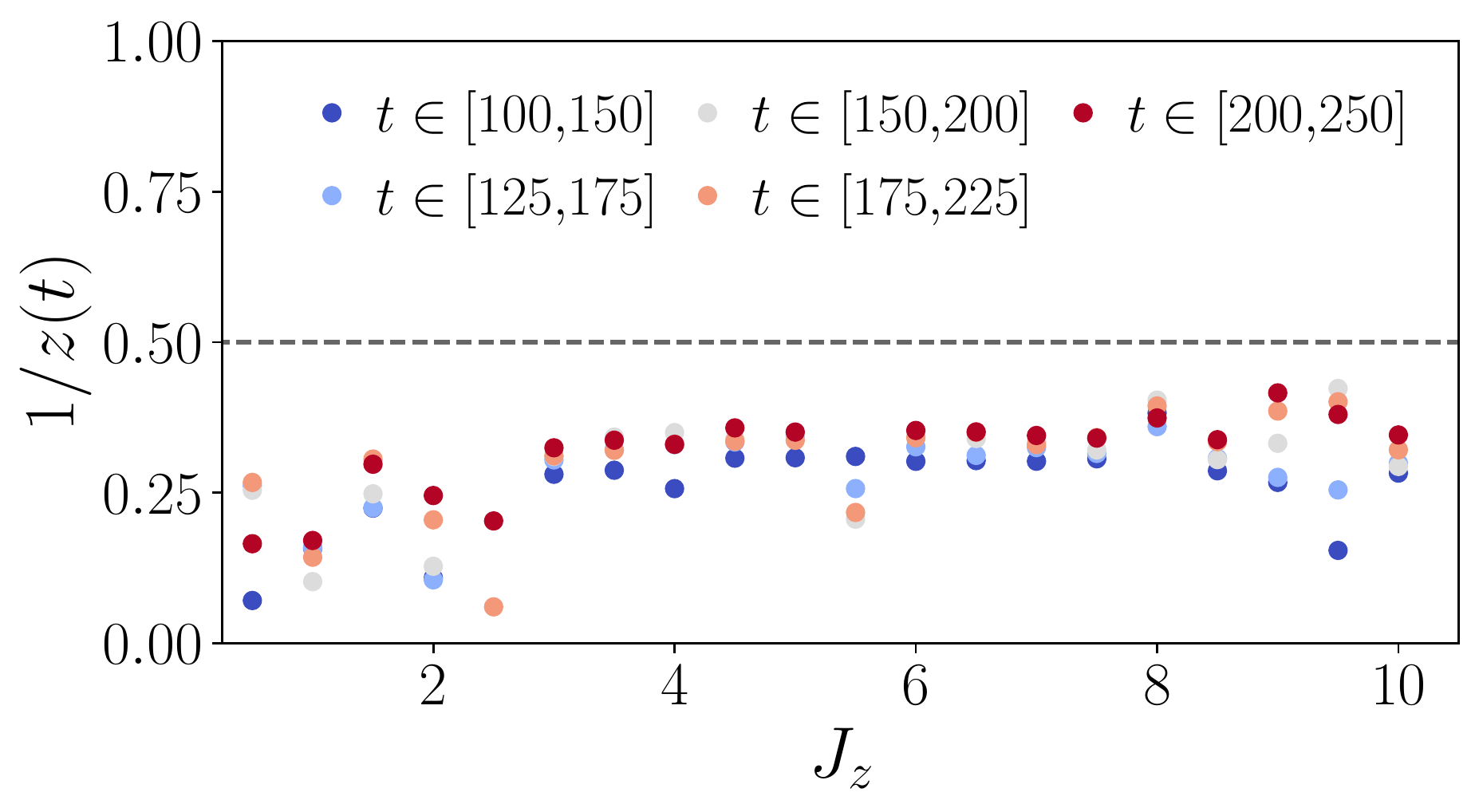}
    \caption{\label{Fig:PRACE-App Heisenberg}
    The inverse dynamical exponent as a function of the interaction strength $J_z$.
    At small values of $J_z$, dynamics are extremely slow and oscillations complicate  the estimation of $1/z$ on available timescales.
    For larger $J_z\geq 3$, $1/z$ shows a more regular behavior, highlighting much slower transport in the Heisenberg chain compared to the two species Hubbard model.
    The data are obtained averaging over $50$ disorder realizations and using a bond dimension $\chi=256$.
    }
\end{figure}

\begin{figure*}
    \centering
    \includegraphics[width=.95\textwidth]{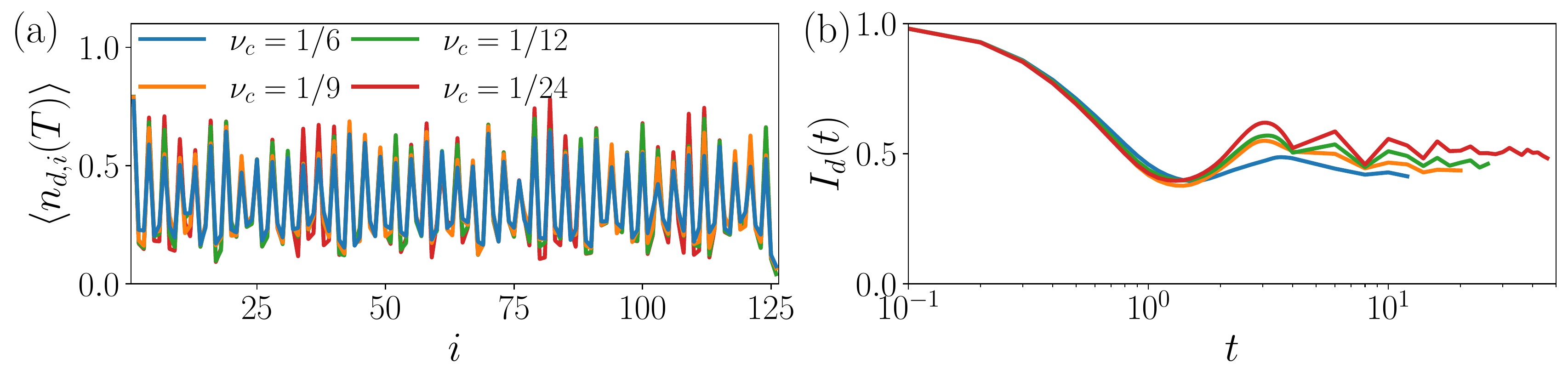}\\
    \includegraphics[width=.95\textwidth]{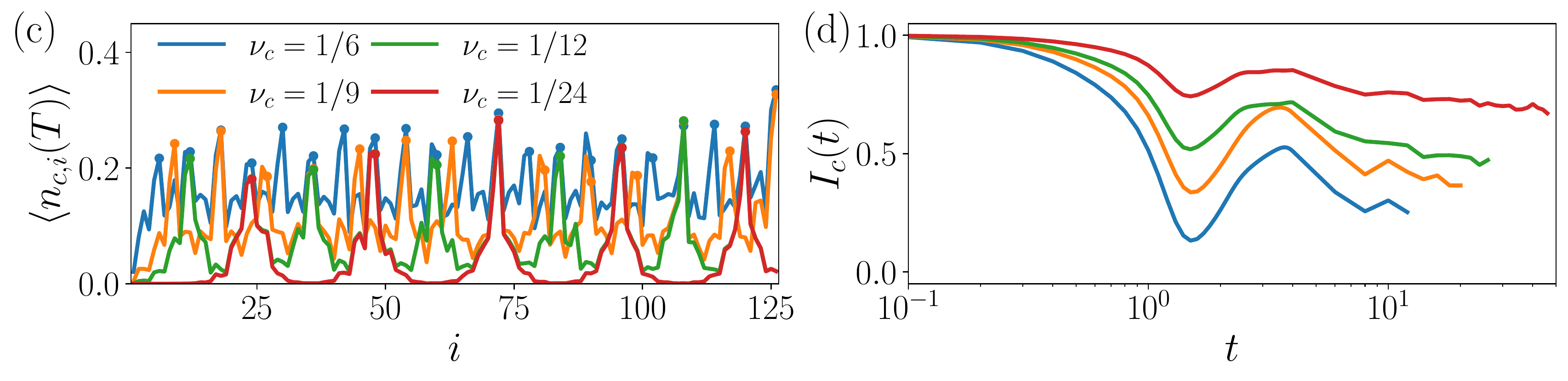}
    \caption{
    The density profiles of both particle types show a clear trend suggesting thermalization at large bath densities.
    However, at $\nu_c=1/24$ $\langle \hat{n}_{c,i}\rangle$ keeps the characteristic density wave structure and the $d$-bosons density profile is significantly farther from relaxation than at larger densities.
    The imbalance confirms a monotonous slow down of the relaxation towards thermal equilibrium as the bath density is decreased.
    The density profiles are shown at the latest time reached by the $\nu_c=1/6$ simulation, the most expensive to simulate, $T=16$. The value of the coupling is fixed to $U=6$.
    }
    \label{Fig:PRACE-App nd nuc}
\end{figure*}

Although the two species Hubbard model and Heisenberg chain have different local Hilbert spaces, one can transform the Hubbard model to spin language to identify the comparable range of parameters.  Under this transformation the hopping $t_{d}$ becomes equivalent to exchange terms that are proportional to $J$ in the Heisenberg model, and we set $J=1$.
The random chemical potential $\epsilon_i$ acting on bosons translates to a random magnetic field $h_i=\epsilon_i/2$. Thus we choose the Heisenberg model with the following parameters, 
\begin{equation}
    \label{Eq:PRACE-App Heisenberg}
\hat{H}=\sum_i \left[s_i^xs_{i+1}^x+s_i^ys_{i+1}^y+\frac{J_z}{2} s_i^zs_{i+1}^z+\frac{\epsilon_i}{2}s_i^z\right].
\end{equation}
We set $W=6.5$ that controls the distribution of $\epsilon_i\in[-W,W]$, and sweep through different interaction strengths $J_z$ since nearest neighbor interaction does not have a direct analog to the on-site Hubbard interaction.
The Heisenberg model~(\ref{Eq:PRACE-App Heisenberg}) conserves total magnetization, allowing for the study of spin transport across a small step, in analogy with the mixed state used in Eq.~(\ref{Eq:rho0}) in the main text.

A similar analysis to the one carried out in the main text results in the inverse dynamical exponent $1/z(t)$ shown in Figure~\ref{Fig:PRACE-App Heisenberg}.
The data reveal much slower transport as compared to the results of Figure~\ref{Fig:zdzc} in the whole parameter range explored, suggesting that the effect of the $c$-bosons cannot be accounted for by an emergent local interaction term among disordered bosons. 
A possible explanation, then, is that the $c$-bosons effectively act as a long-range interaction, justifying the faster transport observed in the two particle species Hubbard model.

\subsection{Dynamics at small densities of clean bosons}\label{Sec:PRACE-App finite nuc}

In the main text, we reported the deviation from logarithmic entanglement growth as a probe of delocalization at small, albeit extensive, bath size.
Here we present some additional data regarding the density profiles and the imbalance, as defined in Eq.~(\ref{Eq:PRACE-App I}).

Figure~\ref{Fig:PRACE-App nd nuc}~(a) and (b) show the behavior of $d$-bosons.
In panel (a) we compare the density profile for different bath densities $\nu_c\in[1/24,1/6]$ at fixed $U=6$ and $W=6.5$.
The relaxation of the initial density wave towards equilibrium becomes monotonously more pronounced as the density, $\nu_c$, is increased, in agreement with the results of the main text.
A similar phenomenology can be observed in the dynamics of the imbalance, which shows much faster decay at $\nu_c=1/6$ compared to  $\nu_c=1/24$.

The bottom panels in Fig.~\ref{Fig:PRACE-App nd nuc} are dedicated to the $c$-bosons.
Similarly to the disordered particles, the density wave is substantially smeared at large $c$-bosons densities, while at $\nu_c=1/24$ large regions with nearly zero density of clean bosons are visible.
Panel~(d) reveals that at every bath size explored in this work the imbalance of the $c$-bosons ($I_c$) decays in time, although in a slower fashion at smaller bath densities.

\begin{figure}[b!]
\includegraphics[width=.99\columnwidth]{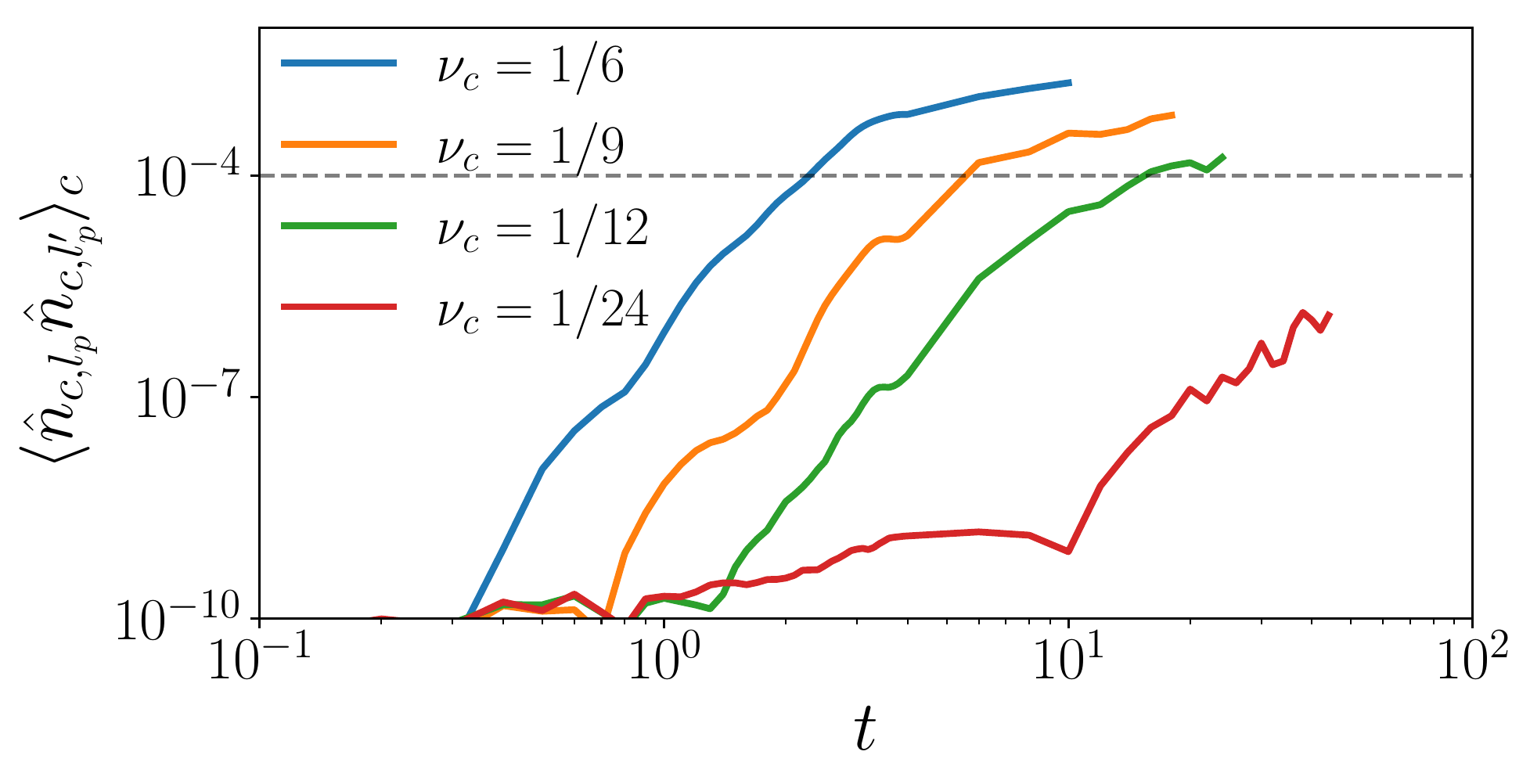}
\caption{\label{Fig:PRACE-App corrc nuc}
Density-density connected correlations among different $c$-bosons at fixed $U=6$. The characteristic time at which they become larger than $\varepsilon = 10^{-4}$ defines an additional timescale $\tau_{cc}(\nu_c)$ that shows approximate power-law dependence on $\nu_c$, $\tau_{cc}\propto 1/\nu_c^k$ with $k\approx -2.53$ for larger values of $\nu_c$.
For the smallest  density $\nu_c=1/24$, correlations may be speculated to show  signatures of saturation to a value much smaller than $\varepsilon$, but longer times are needed to verify this hypothesis.
}
\end{figure}

Finally, we analyze the connected correlation function of  $c$-bosons. 
The data shown in the main text leads to the hypothesis that the deviation from logarithmic growth of the entanglement entropy could be generated by a significant correlation among the different $c$-bosons.
To check this hypothesis, we studied the behavior of the density-density connected correlation function among different sites $l_p$ and $l^{'}_{p}$, where distance between sites $l_p$ and $l_p'$ corresponds to the initial position of adjacent $c$-bosons.
The results are shown in Figure~\ref{Fig:PRACE-App corrc nuc}.
The density-density correlations start growing at progressively earlier times as the density $\nu_c$ is increased and the correlation function crosses the threshold value $\varepsilon=10^{-4}$ (dashed line) at a timescale $\tau_{cc}$ scaling in bath size with the same power-law observed for the onset of the deviation of entanglement growth from logarithmic. This indeed suggests that correlations among clean bosons may be responsible for the onset of more rapid entanglement growth shown in Fig.~\ref{Fig:S nuc} in the main text.
Also, consistent with all other probes, the connected correlation function for the lowest density case, $\nu_c=1/24$, does not cross the threshold value $\varepsilon$ within the available simulation time. 

%

\end{document}